\documentclass[preprint]{JHEP3}

\usepackage{amsmath,amssymb,epsfig}

\newcommand{\beq}{\begin{equation}} 
\newcommand{\eeq}{\end{equation}\noindent}
\newcommand{\bean}{\begin{eqnarray*}}
\newcommand{\eean}{\end{eqnarray*}\noindent}
\newcommand{\bea}{\begin{eqnarray}}
\newcommand{\eea}{\end{eqnarray}\noindent}

\newcommand{\bx}{{\vec x}}
\newcommand{\by}{{\vec y}}

\newcommand{\bp}{{\vec p}}
\newcommand{\bq}{{\vec q}}
\newcommand{\bk}{{\vec k}}

\newcommand{\bPi}{\bar{\Pi}}
\newcommand{\bG}{\bar{G}}
\newcommand{\bcG}{\bar{\delta\cal G}}
\newcommand{\bF}{\bar{\cal F}}
\newcommand{\PP}{{\cal P}}

\newcommand{\pol}{{(\lambda)}}
\newcommand{\dl}{ {\ell^+\ell^-} }
\newcommand{\G}{ \delta{\cal G} }
\newcommand{\D}{ {\cal D} }
\newcommand{\Op}{ {\cal O} }
\newcommand{\C}{ {\cal C} }
\newcommand{\Sp}{ {\cal S} }
\newcommand{\e}{ {\rm e} }
\newcommand{\emg}{ {\rm em} }
\newcommand{\dd}{ {\rm d} }
\newcommand{\s}{ {\rm s} }
\newcommand{\I}{ {\rm I} }
\newcommand{\T}{ {\rm T} }
\newcommand{\tr}{{\rm tr}}
\newcommand{\Tr}{{\rm Tr}}
\renewcommand{\Re}{{\rm Re}}
\renewcommand{\Im}{{\rm Im}}
\renewcommand{\slash}{\!\!\!\!/\,}

\preprint{HD--THEP--03--48}

\title{Out-of-equilibrium electromagnetic radiation}

\author{Julien Serreau\\
	Institut f\"ur Theoretische Physik der Universit\"at Heidelberg\\
	Philosophenweg 16, D-69120 Heidelberg, Germany\\
	E-mail: \email{serreau@thphys.uni-heidelberg.de}}

\abstract{
We derive general formulas for photon and dilepton production rates
from an arbitrary non-equilibrated medium from first principles in 
quantum field theory. At lowest order in the electromagnetic coupling 
constant, these relate the rates to the unequal-time in-medium photon 
polarization tensor and generalize the corresponding expressions for 
a system in thermodynamic equilibrium. We formulate the question of 
electromagnetic radiation in real time as an initial value problem 
and consistently describe the virtual electromagnetic dressing of 
the initial state. In the limit of slowly evolving systems, we recover
known expressions for the emission rates and work out the first
correction to the static formulas in a systematic gradient expansion.
Finally, we discuss the possible application of recently developed 
techniques in non-equilibrium quantum field theory to the problem of 
electromagnetic radiation. We argue, in particular, that the 
two-particle-irreducible (2PI) effective action formalism provides 
a powerful resummation scheme for the description of multiple 
scattering effects, such as the Landau-Pomeranchuk-Migdal 
suppression recently discussed in the context of equilibrium QCD.
}

\begin{document}

\section{Introduction}
\subsection{Motivations and overview of previous works}

Because they interact weakly with matter, direct electromagnetic signals 
provide sensitive probes of the hot and dense matter produced in high 
energy heavy ion collisions~\cite{Feinberg:1976ua}. 
Once emitted, they mostly escape the collision zone without further 
interaction and, therefore, carry direct information about the state 
of the emitting system at the various stages of its time 
evolution. Recently, the WA98 collaboration has reported the first 
observation of direct photons in central $Pb$--$Pb$ collisions at the 
CERN SPS~\cite{Aggarwal:2000th}. The measurement of direct photons and 
dileptons is one of the major goals of the PHENIX experiment at RHIC 
\cite{Morrison:1998qu} and is among the key observables to be studied 
by the ALICE collaboration at the LHC~\cite{Alessandro:bv}.

Electromagnetic probes are expected to be directly sensitive to the 
(partonic or hadro\-nic) nature of the relevant degrees of freedom 
underlying the dynamics of the produced matter. They have been proposed 
as a promising signature of the possible formation of a deconfined 
state of matter, the so-called quark-gluon 
plasma~\cite{Shuryak:1978ij,Kajantie:wg}. An intense theoretical 
activity has been devoted to study the emissive power of a 
quark-gluon plasma vs. that of a hot hadron gas in thermodynamic 
equilibrium (for recent reviews see~\cite{Peitzmann:2001mz,Gale:2003iz}). 
Both are highly non-trivial problems, which require, in particular, a 
detailed understanding of various medium effects such as, for instance 
in the former case, dynamical screening~\cite{Kapusta:qp,Baier:1991em}, 
or interference effects between multiple scattering processes, 
the so-called Landau-Pomeranchuk-Migdal
effect~\cite{Aurenche:2000gf,Arnold:2001ba,Aurenche:2002wq} 
(for a recent short review, see~\cite{Gelis:2002yw}). 
To make contact with phenomenology, one has to model the space-time 
evolution of the fireball. For this purpose, one usually makes the 
simplifying assumption that, due to their multiple interactions, the 
produced partons thermalize (locally) shortly after the initial impact. 
The integration over space-time is then performed by folding the 
equilibrium rates, computed in either the deconfined or the hadronic 
phase, with a given hydrodynamic model (see e.g.\ \cite{Steffen:2001pv}), 
which includes, in particular, a modelization of the equation of state 
in both phases as well as of the phase transition (for a recent review,
see~\cite{Kolb:2003dz}).

However, although ideal hydrodynamics is doing remarkably well in 
describing data such as elliptic flow or transverse spectra at not too high 
momenta~\cite{Heinz:2001xi}, the 
assumption of early thermalization appears questionable, at least from 
the point of view of perturbative QCD (see e.g.\ \cite{Mueller:2002kw}).
Indeed, detailed theoretical studies of the early-time evolution of the 
partonic system reveal that the so-called pre-equilibrium stage may not
be as short as usually believed. In particular, in very energetic 
collisions at RHIC and at LHC, where it is expected that the partons 
produced in the central rapidity region are mostly gluons, it appears 
that elementary QCD processes are not fast enough to lead to efficient 
chemical equilibration between gluon and (anti-)quark 
populations~\cite{Biro:1993qt}. 
Moreover, recent studies of kinetic equilibration have shown that it is 
extremely difficult to reach thermal equilibrium on time scales lower 
than a few fm/c, due to the strong longitudinal expansion at very early 
times~\cite{Baier:2002bt}. 
Although the issue of early thermalization is still a
matter of debate (see e.g.~\cite{Serreau:2002yr}), it seems important 
to consider the influence of the non-equilibrium evolution on the 
various experimental signatures of the produced matter and, in 
particular, on electromagnetic signals, which are sensitive to 
the earliest stages of the collision.

There are various other motivations for studying out-of-equilibrium 
electromagnetic radiation in the context of heavy ion collisions. For 
instance, it is expected that a non-trivial semi-classical pion field 
configuration might form during the rapid cooling of the system 
through the chiral phase transition~\cite{Rajagopal:1993ah}. The 
electromagnetic radiation associated with the decay of this so-called 
disoriented chiral condensate has been proposed as a possible signature 
of the phenomenon~\cite{Huang:1996kq,Boyanovsky:1997cy}, which in turn
might provide a useful tool to locate the critical point 
of the QCD phase diagram~\cite{Rajagopal:2000yt}. Another interesting 
motivation is related to the fact that, because of their sensitivity
to the early stages of the collision, direct electromagnetic signals
might bring experimental information concerning the initial conditions 
realized after the nuclear impact.

Various aspects of out-of-equilibrium electromagnetic radiation
have been addressed in the literature, in particular in the context 
of an under-saturated partonic plasma, where local kinetic equilibrium is 
assumed to be rapidly achieved, but where quarks and gluons are not yet
chemically equilibrated~\cite{Shuryak:1992bt,Traxler:1995kx,Baier:1997xc}. 
In that case, the momentum distributions of quarks and gluons are isotropic 
and one can closely follow equilibrium calculations. In particular, one can 
generalize the so-called hard thermal loop (HTL) resummation 
and extend the usual analysis of dynamical screening to
the non-equilibrium case~\cite{Baier:1997xc}.
So far, these calculations have been pushed up to the two-loop 
level~\cite{Dutta:2001ii} and lead to significantly different results from 
the corresponding equilibrium calculations because of the low population of 
quarks and anti-quarks at early times~\cite{Mustafa:2000sg,Dutta:2001ii}. 
The case of a system out of kinetic equilibrium has been much less studied.
Attempts in this direction generally rely on folding general distribution 
functions, taken to be the solutions of appropriate Boltzmann equations, 
with conventional matrix elements~\cite{Wong:1998tr,Bass:2002pm}. 
This, however, does not provide a consistent description of phenomena like
screening and one generally has to introduce an infra-red cut-off by hand. 
Although one can also formulate the analog of the HTL resummation in that 
case (see e.g.\ \cite{Carrington:1997sq}), it is more difficult to employ 
in practice because momentum integrations cannot be explicitly performed 
for arbitrary distribution functions. 

An important remark to be done concerning the above calculations is that,
unlike their equilibrium counterpart, they do not rely on a first principle 
derivation of the out-of-equilibrium photon and dilepton production 
rates. For instance, the most widely used ansatz is to start with 
equilibrium-like expressions for the rates, written at a given space-time
point, and to simply replace the Bose-Einstein and Fermi-Dirac distribution 
functions appearing e.g.\ in the real-time equilibrium formalism by other 
functions of momentum reflecting the out-of-equilibrium particle 
distribution in phase-space. This procedure would be exact for general
stationary systems; it might also be justified when the duration of the 
emission process is short enough as compared to the typical relaxation 
time of the emitting system~\cite{Gelis:2001xt}; but it can clearly not
be universally valid for computing non-equilibrium processes. In the 
general case, one expects the emission process to depend non-locally on
the space-time history of the emitting system and the main issue is to properly
formulate the production rate itself. An attempt in this direction has been
made in Ref.\ \cite{Niegawa:1999jr}. However, the approach presented there 
is widely based on the use of standard perturbation theory, which is known 
to be badly suited for the description of non-equilibrium 
systems.\footnote{For instance, the 
standard perturbative expansion is known to be poorly convergent out
of equilibrium, due to the occurrence of spurious pinch singularities
\cite{Altherr:1994fx,Bedaque:1994di}, which lead, in particular, to 
so-called secular terms~\cite{Greiner:1998ri,Boyanovsky:1998tg}
(these singularities cancel out in thermodynamic equilibrium 
thanks to detailed balance relations~\cite{Baier:1988xv,LeBellac:book}).}
Interesting steps towards a more satisfying formulation of the problem 
have been made recently in a slightly different context in Refs.\ 
\cite{Wang:2000pv,Moore:CERN,Boyanovsky:2003qm}, where the authors
discuss possible finite lifetime effects for a system in equilibrium. 
Still, the real-time formulation of \cite{Wang:2000pv} suffers from 
serious physical problems such as spontaneous photon emission from the 
vaccum or infinite amounts of radiated energy \cite{Moore:CERN}. 
These difficulties can be traced back to the issue of properly 
identifying and subtracting the unobservable virtual radiation at 
any finite time \cite{Boyanovsky:2003qm}.

\subsection{Summary of the present work}

In this paper, we address the question of out-of-equilibrium
electromagnetic radiation from a very general point of view and 
present a first principle derivation of photon and dilepton emission 
rates in the framework of non-equilibrium quantum field theory.
We treat electromagnetic interactions at lowest order in perturbation 
theory but, otherwise, make no assumption concerning the internal 
dynamics of the emitting system. Our approach generalizes the real-time 
formulation of Boyanovsky {\it et al.} \cite{Wang:2000pv} to the case 
of arbitrarily far-from-equilibrium systems. In order to deal with the 
problem of virtual radiation mentioned above, we propose to consistently
take into account the electromagnetic dressing of the initial state. In 
particular, this allows one to unambiguously identify the virtual contribution 
at the initial time and to subtract it at once.

We obtain general formulas relating the out-of-equilibrium production rates
to the intrinsic dynamics of the emitting system which, at lowest order in the 
electromagnetic coupling constant, is characterized by the non-equilibrium 
in-medium photon polarization tensor. This generalizes the corresponding 
formulas in equilibrium~\cite{Feinberg:1976ua,Gale:1990pn,McLerran:1984ay}, 
which we recover in the appropriate limit. In particular, we show that the 
description of the initial virtual cloud is essential in order to recover 
known formulas for the particular case of stationary systems. We also discuss 
the case of slowly evolving (quasi-stationary) systems and derive the first 
corrections to the static photon and dilepton production rates in a standard 
gradient expansion. 

In the last part of the paper, we make contact with recent progress 
concerning the description of far-from-equilibrium quantum fields, 
based e.g.\ on the so-called two-particle-irreducible (2PI) effective 
action formalism \cite{Berges:2000ur}, or on truncations of Schwinger-Dyson 
equations \cite{Mihaila:2000sr} (for a recent review, see
\cite{Berges:2003pc}). In particular, these methods allow for practical, 
first principle calculations of the non-equilibrium dynamics in realistic 
particle physics applications \cite{Berges:2002cz,Berges:2002wr}. We discuss 
how they might be employed in the present context and propose a formalism, 
based on introducing a $\Sp$--shape path in real time, which allows one to 
generalize existing methods to the case of dressed initial states. Finally, 
we outline how the relevant current-current correlator can be obtained 
within the 2PI effective action formalism. We point out that finite order
approximations of the 2PI effective action automatically generates infinite
resumations of ladder-type diagrams, which correspond to multiple scattering 
processes. Recall that the latter have recently been shown to play a crucial 
role in obtaining the full photon and dilepton production rates from an equilibrated
QCD plasma at leading order in $\alpha_s$ \cite{Arnold:2001ba,Aurenche:2002wq}.

\section{The non-equilibrium set-up}
\label{sec:noneq}

We consider a system of interacting (bosonic and/or fermionic) fields,
prepared in a given out-of-equilibrium initial state, and which emits 
photons and lepton pairs in the course of its time evolution. 
We assume that the relative energy loss is negligible so that the
process of electromagnetic emission can be considered as a small 
perturbation. Our aim in this section is to relate, at lowest order in 
perturbation theory, the inclusive spectrum of radiated photons and 
lepton pairs to the intrinsic non-equilibrium dynamics of the emitting 
system, without further approximations. 

\subsection{Initial value problem: Schwinger's closed time-path}

We denote by $H_\s$ the Hamiltonian describing the internal dynamics of
the emitting system which we assume has been prepared at time 
$t=0$ in an arbitrary state characterized by the (normalized) density 
matrix $\rho_\s$. In terms of the eigenstates $|n,\s\rangle$ of $H_\s$,
where $n$ denotes a set of relevant quantum numbers, the latter can
be written
\beq
\label{dmsyst}
 \rho_\s = \sum_{n,m} P^\s_{nm}\, |n,\s\rangle\langle m,\s| \,,
\eeq
with $P^\s_{nm}\equiv\langle n,\s|\rho_\s |m,\s\rangle$.
Similarly, we denote by $H_\emg$ the Hamiltonian describing
free propagation of electromagnetic degrees of freedom, that is photons 
and (anti-)leptons, and by $\rho_\emg$ the density matrix characterizing
the initial state of the electromagnetic sector. As above, one can 
write:
\beq
\label{dmemg}
 \rho_\emg = \sum_{a,b} P^\emg_{ab}\, |a,\emg\rangle\langle b,\emg| 
\eeq
in terms of the eigenstates $|a,\emg\rangle$ of $H_\emg$.

In absence of electromagnetic interactions, the most general density 
matrix for the whole system `emitter+radiation' has the factorized form:
\beq
\label{factorized}
 \bar{\rho} = \rho_\s \otimes \rho_\emg\,.
\eeq
At this order of approximation, the two subsystems evolve independently: 
The time-evolution of the emitter is driven by the interactions
between its internal degrees of freedom, whereas the (free) electromagnetic 
sector undergo a trivial evolution from the initial state (\ref{dmemg}).
Introducing electromagnetic interactions, the total Hamiltonian reads:
\beq
\label{Hamtot}
 H = H_\s + H_\emg + H_{\rm int} \, ,
\eeq
where $H_{\rm int}$ represents the coupling of the photon field with the 
degrees of freedom of the emitting system as well as with leptons. 
In general, electromagnetic interactions might also introduce non-trivial 
correlations between the emitting system and the electromagnetic degrees
of freedom in the initial state and the total density matrix $\rho$ cannot 
be factorized as in Eq.\ (\ref{factorized}). As mentioned in the 
introduction, we are interested in describing the virtual electromagnetic 
cloud associated with the charged degrees of freedom of the emitting system. 
The corresponding density matrix can be formally obtained by ``dressing'' the bare 
initial state (\ref{factorized}): The dressed states $|n,a\rangle$ can be 
expressed in terms of the unperturbed states as follows (up to an overall 
phase)~\cite{Bohm:book} (see also \cite{Boyanovsky:2003qm}):
\beq
\label{adiab}
 |n,a\rangle=U_\epsilon (0,-\infty)\,|n,\s\rangle\otimes |a,\emg\rangle\,,
\eeq
where the unitary ``dressing'' operator $U_\epsilon (0,-\infty)$ is
formally equivalent to an adiabatic switching-on of electromagnetic 
interactions from the infinite past:
\beq
\label{dressing}
 U_\epsilon  (0,-\infty)= \T \exp \Big\{-i\int_{-\infty}^{0} du \, 
 \e^{\epsilon u} \, H_\I (u) \Big\} = U^\dagger_\epsilon  (-\infty,0) \, .
\eeq
with $\epsilon \equiv 0^+$.\footnote{For the leading-order perturbative
treatment presented below, it is actually sufficient that $\epsilon/\bar{k}\ll e$,
where $\bar{k}$ is a typical energy scale in the problem.}
\beq
 H_\I (t) = 
 \e^{i(H_\s+H_\emg)t} \, H_{\rm int} \, \e^{-i(H_\s+H_\emg)t} \, .
\eeq
Using Eqs.\ (\ref{dmsyst})-(\ref{factorized}) and (\ref{adiab}), one obtains 
the required dressed density matrix, in terms of the bare one: 
\beq
\label{dmatrix}
 \rho = U_\epsilon (0,-\infty) \, \bar{\rho} \, U_\epsilon^\dagger (0,-\infty) \, .
\eeq

\begin{figure}[t]
 \begin{center}
 \epsfig{file=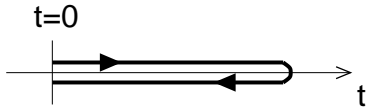,width=4.8cm}
 \end{center}
 \caption{\label{fig:CTP}
 The Schwinger closed time path. In this formulation, the initial condition 
 is specified at the initial time $t=0$ and is characterized by the dressed
 density matrix $\rho$.}
\end{figure}

Once the initial state has been specified, the subsequent non-equilibrium 
evolution is completely determined by the full Hamiltonian $H$.
We shall typically be interested in unequal-time correlation functions 
of the form:
\beq
\label{correlfunc}
 \langle \Op (t_1) \cdots \Op (t_n) \rangle \equiv
 \Tr [\,\rho\,\Op (t_1) \cdots \Op (t_n)] \, ,
\eeq
where the time evolution of a given operator $\Op$ is given by
\beq
  \Op (t) = \e^{iH\,t}\,\Op\,\e^{-iH \,t} \, .
\eeq
For given initial conditions, specified at $t=0$ and characterized by 
the density matrix $\rho$, the above correlation functions may be 
computed by introducing a finite closed path in real time~\cite{Schwinger:1960qe}, 
as shown in Fig.\ \ref{fig:CTP}. 

\subsection{Virtual cloud vs. virtual time evolution: revisiting Keldysh's time-path}

In this subsection, we exploit the preceeding adiabatic construction, 
in order to replace the initial correlations due to electromagnetic dressing 
by a formally equivalent time evolution from an uncorrelated state of the form
(\ref{factorized}) in the infinite past. This allows us to treat the 
virtual dressing of the initial state and the process of radiation
during the physical time evolution on a similar footing and considerably
simplifies the calculations.\footnote{We mention that a similar procedure 
is employed in describing interacting fields in thermodynamic equilibrium. 
There, the non-Gaussian density matrix is dealt with by introducing a formal
``time evolution'' in the imaginary-time direction.} 

\begin{figure}[t]
 \begin{center}
 \epsfig{file=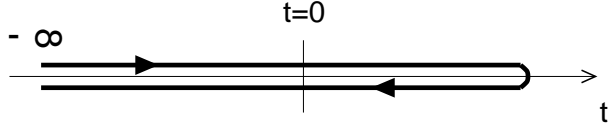,width=8.cm}
 \end{center}
 \caption{\label{fig:Kpath}
 The electromagnetic dressing of the initial state at time $t=0$
 is formally equivalent to a time evolution from a state characterized
 by the bare density matrix $\bar{\rho}$ at $t=-\infty$ to the physical
 initial time $t=0$. This can be described by introducing the above (Keldysh)
 path in real time.}
\end{figure}

To make this more precise, we make use of the interaction picture 
with respect to the unperturbed Hamiltonian $H_\s+H_\emg$ introduced 
above:
\beq
\label{intevol}
 \Op_\I (t) = 
 \e^{i(H_\s+H_\emg)t} \, \Op \, \e^{-i(H_\s+H_\emg)t} \, .
\eeq
Interaction picture operators are related to Heisenberg operators
via:
\beq
\label{int1}
 \Op (t) = U^\dagger(t,0)\,\Op_I (t)\,U(t,0)\, ,
\eeq
where
\beq
\label{int2}
 U(t,t')=\T \exp \Big\{-i\int_{t'}^t du \, H_\I (u) \Big\}
 = U^\dagger (t',t)\, .
\eeq
For the sake of the argument, we focus here on the case of a two-point 
correlation function. The generalization to higher correlation functions
is straightforward. Using the form (\ref{dmatrix}) of the density matrix and 
noticing that 
\beq
 \Tr [\rho]=\Tr [\bar{\rho}]=1\, ,
\eeq
we can write ($t>t'\ge 0$):
\bea
\label{reduc}
 \langle \Op (t)\,\Op (t')\rangle
 &=&\Tr \Big[\,\rho\,\,
 U(0,t)\,\Op_\I (t)\,U(t,t')\,\Op_\I (t')\,U(t',0)\Big]\nonumber\\
 &=&\Tr \Big[\,\bar{\rho}\,\,
 U_\epsilon (-\infty,0)\,U(0,t)\,
 \Op_\I (t)\,U(t,0)\,U_\epsilon (0,-\infty)\nonumber\\
 &&\quad\times U_\epsilon (-\infty,0)\,U(0,t')\,
 \Op_\I (t')\,U(t',0)\,U_\epsilon (0,-\infty) \Big]\nonumber\\
 &=&\Tr \Big[\,\bar{\rho}\,\,
 V_\epsilon(-\infty,t)\,\Op_\I (t)\,V_\epsilon(t,t')\,
 \Op_\I (t')\,V_\epsilon(t',-\infty)\Big]\,,
\eea
where we defined the unitary operator 
\beq
 V_\epsilon (t,t')= \T\exp\Big(-i\int_{t'}^t[du]\,H_\I (u)\Big) 
 = V^\dagger_\epsilon(t',t)\, ,
\eeq
which satisfies the relation:
\beq
 V_\epsilon (t,-\infty) = U(t,0)\,U_\epsilon (0,-\infty)\,.
\eeq
Here, we introduced the shorthand notation: 
\beq
\label{timeint}
 [du] \equiv du\,[\,\theta(-u)\,\e^{\epsilon u}+\theta(u)\,] \, .
\eeq
Reading the first line of Eq.\ (\ref{reduc}) from right to left, one
clearly sees the time evolution from $t=0$ to the positive time $t'$, then 
to $t>t'$ and, finally, back to $t=0$. This is the essence of the 
closed time path formulation corresponding to the contour of 
Fig.\ \ref{fig:CTP}. Here, the initial state is specified at $t=0$
and is characterized by the dressed density matrix $\rho$. 
Similarly, we clearly see, from the third line of Eq.\ (\ref{reduc}), 
that this can be reformulated as a ``time-evolution'' along the 
contour of Fig.\ \ref{fig:Kpath}, which extend to the infinite past, 
and where the ``initial'' state, now specified at $t=-\infty$, is
characterized by the bare density matrix $\bar{\rho}$. The closed path
$\C$ along the real axis, represented on Fig.\ \ref{fig:Kpath}, is 
sometimes referred to as the Keldysh contour~\cite{Keldysh:ud} in the 
literature. It is useful to introduce time-ordered correlation functions 
along the contour, which can be written as (see Eq.\ (\ref{reduc})):
\beq
 \Big< T_\C\Big\{\Op (t_1)\cdots\Op (t_n)\Big\}\Big>
 =\Tr \left[\,\bar{\rho}\,\,T_\C\Big\{\Op_\I (t_1)\cdots\Op_\I (t_n)\,
 \exp\Big(-i\int_\C[du]\,H_\I (u)\Big)\Big\}\right]\,,
\eeq
where $T_\C$ represents time-ordering along $\C$. We mention that one
can equivalently formulate the above construction in terms of path 
integrals by means of standard manipulations (see also section 
\ref{sec:polar} below).

In the following, we shall use standard notations for two-point 
correlation functions:
\bea
\label{Tcorrelator}
 {\rm C}_{\alpha\beta} (t,t') &\equiv&
 \Big< \T_\C \Big\{ \Op_\alpha (t)\,\Op_\beta(t') \Big\} \Big>\nonumber\\
 &=&\Theta_\C (t-t')\,{\rm C}^>_{\alpha\beta} (t,t') \pm 
 \Theta_\C (t'-t)\,{\rm C}^<_{\alpha\beta} (t,t') \, ,
\eea
where $\Op_\alpha$ and $\Op_\beta$ are two arbitrary operators and 
$\alpha$ and $\beta$ represent a set of (Dirac, Lorentz, etc \ldots) 
indices and/or spatial variables. The plus/minus sign corresponds 
to the case of bosonic/fermionic operators respectively and 
the the two independent components are given by:
\bea
\label{correlsup}
 {\rm C}^>_{\alpha\beta} (t,t')&=&
 \langle\Op_\alpha (t)\,\Op_\beta(t')\rangle\, ,\\
\label{correlinf}
 {\rm C}^<_{\alpha\beta} (t,t')&=&
 \langle \Op_\beta(t')\,\Op_\alpha (t)\rangle\, .
\eea

\section{Out-of-equilibrium electromagnetic radiation}
\label{sec:emrad}

We now apply the formalism described above to the calculation of the 
out-of-equilibrium photon and dilepton production rates as a function 
of time, at leading order in the electromagnetic coupling constant. 
Assuming that no real electromagnetic radiation is present initially, 
the initial state is described by Eqs.\ (\ref{factorized}) and 
(\ref{dmatrix}) with
\beq
\label{gammavac}
 \rho_\emg = |0,\emg \rangle\langle0,\emg| \, .
\eeq
The generalization of the following considerations to the case where real 
radiation is already present in the initial state is straightforward. Note 
that the density matrix $\rho_\s$ describing the initial state of the emitting 
system is left arbitrary.
We assume that the electromagnetic interaction Hamiltonian has the 
following form (in the Schr\"odinger picture):
\beq
\label{JA}
 H_{\rm int} = \int d^3 x\,{\cal J}^\mu (\bx)\,A_\mu (\bx) \, ,
\eeq
where $A^\mu (\bx)$ is the electromagnetic field operator and 
\beq
 {\cal J}^\mu (\bx)= J^\mu (\bx) + j^\mu (\bx)
\eeq
is the total electromagnetic current, including the contributions 
from both the emitting system ($J^\mu$) and the leptonic fields 
$\psi$:
\beq
\label{emcurrent}
 j^\mu (\bx)=e\,\bar{\psi} (\bx) \gamma^\mu\psi (\bx)\,.
\eeq
Here, $e$ denotes the electromagnetic coupling constant and $\gamma^\mu$
are the usual Dirac matrices. In the case where the emitting system 
contains charged scalar fields, there is an additional contribution 
-- quadratic in the photon field -- to the interaction Hamiltonian 
(\ref{JA}). We shall consider such a term at the end of this section.
From now on, we specialize to the case of spatially homogeneous systems
for simplicity.

\subsection{Inclusive photon spectrum at time $t$}

At a given time $t$, the single-particle inclusive photon spectrum can be 
expressed as:\footnote{Strictly speaking, the following expression (see 
also Eq.\ (\ref{llspectrum}) below) represents the spectrum of photons
present in the system at time $t$. This can be identified with the spectrum 
of photons which have been emitted after a time $t$ under the assumption 
that produced electromagnetic radiation escape the system without further 
interaction. The physically relevant limit corresponds to $t\rightarrow+\infty$.}
\beq
\label{gspectrum}
 \frac{dn_\gamma (t,\bk)}{d^3x\,d^3k} = \frac{1}{(2\pi)^3} \, 
 \sum_{\lambda=\perp} \Big< N^\pol_\gamma (t,\bk) \Big> \, ,
\eeq
where the sum runs over physical (transverse) polarizations. Here, 
\beq
 N^\pol_\gamma (t,\bk)\equiv 
 a_\lambda^\dagger (t,\bk) \, a_\lambda (t,\bk)
\eeq 
denotes the Heisenberg picture of the photon number operator for momentum $\bk$
and polarization $\lambda$. The corresponding creation and annihilation 
operators are related to the electromagnetic field operator and its time
derivative through:
\beq
\label{anihil}
 a_\lambda(t,\bk)=-\frac{\epsilon^\pol(\bk)}{\sqrt{2k}}\cdot
 \Big[i\,\dot{A}(t,\bk)+k\,A(t,\bk)\Big]\,,
\eeq
with appropriate polarization vectors $\epsilon^\pol_\mu(\bk)$.
Here, we defined the Fourier modes of the field as ($V\equiv (2\pi)^3\,
\delta^{(3)} (\vec{0})$ is the normalization volume):
\beq
 A_\mu (t,\bk)=\frac{1}{\sqrt V}\int d^3x\,
 \e^{-i\bk\cdot\bx}\,A_\mu (t,\bx)
\eeq
and similarly for $\dot{A}_\mu (t,\bk)\equiv \partial_t A_\mu(t,\bk)$.
Note that the operators $a(t,\bk)$ and $a^\dagger(t,\bk)$ are to be 
understood as annihilating and creating excitations above the averaged 
(macroscopic) electromagnetic field which can be non-zero in general. 
Therefore, the averaged value in Eq.\ (\ref{gspectrum}) is to be 
understood as a connected correlator: $\langle a^\dagger\,a\rangle_c\equiv 
\langle (a^\dagger-\langle a^\dagger\rangle)(a-\langle a\rangle)\rangle=
\langle a^\dagger\,a\rangle-\langle a^\dagger\rangle\langle a \rangle$.
All correlation functions we shall encounter in the following denote 
connected correlators as well and we omit the subscript `$c$' for 
convenience.
Using Eq.\ (\ref{anihil}), we get
\beq
\label{gnumber1}
 \sum_{\lambda=\perp} \Big< N^\pol_\gamma (t,\bk) \Big> = 
 \frac{\partial_t\partial_{t'}+ik(\partial_t-\partial_{t'})+k^2}{2k}\,
 \gamma^{\mu\nu} (\bk)\,{\cal G}^<_{\nu\mu} (t,t';\bk)\Big|_{t'=t}\,, 
\eeq
where we introduced the connected photon propagator:
\beq
\label{gprop}
 {\cal G}_{\mu\nu} (t,t';\bk) \equiv 
 \Big< \T_\C \Big\{ A_\mu (t,\bk)\,A_\nu^\dagger(t',\bk) \Big\} \Big>
\eeq
as well as the photon tensor ($g_{\mu\nu}={\rm diag}(1,-1,-1,-1)$)
\beq
 \gamma_{\mu\nu} (\bk) \equiv 
 \sum_{\lambda=\perp} \epsilon^{\pol *}_\mu (\bk)\,\epsilon^\pol_\nu (\bk) =
 -g_{\mu\nu}-\kappa_\mu \kappa_\nu + n_\mu n_\nu \, .
\eeq
For transverse polarizations, 
$K\cdot\epsilon^\pol\equiv K^\mu\, \epsilon^\pol_\mu=0$, where 
$K^\mu\equiv (k^0,\bk)$ is the four-momentum of the photon. 
Here, we chose the (space-like) transverse polarization vectors to 
be orthonormal: $\epsilon^\pol \cdot\epsilon^{(\lambda')} = 
-\delta^{\lambda\,\lambda'}$, and we introduced the time-like unit 
vector $n^\mu$ ($n^2=1$) such that $n\cdot \epsilon^\pol=0$ as well 
as the space-like unit vector ($\kappa^2=-1$)
\beq
 \kappa^\mu\equiv\frac{K^\mu-(K\cdot n)\,n^\mu}{\sqrt{(K\cdot n)^2-K^2}} \,,
\eeq
which satisfies $\kappa\cdot\epsilon^\pol=\kappa\cdot n= 0$.
For on-shell photons ($K^2=0$), one has
\beq
\label{gtensor}
 \gamma_{\mu\nu} (\bk) = -g_{\mu\nu}
 +\frac{K_\mu n_\nu + K_\nu n_\mu}{K\cdot n}
 -\frac{K_\mu K_\nu}{(K\cdot n)^2} \, .
\eeq

\begin{figure}[t]
 \begin{center}
 \epsfig{file=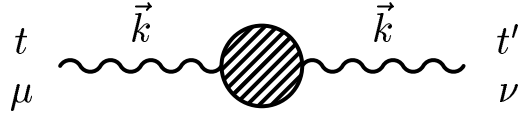,width=7.cm}
 \end{center}
 \caption{\label{fig:gamma1}
 The in-medium contribution $\G_{\mu\nu} (t,t';\bk)$ to the 
 non-equilibrium photon propagator at lowest order in the electromagnetic 
 coupling constant $e$. Wavy lines represent the free vacuum photon 
 propagator and the dashed blob represents the unequal-time current-current 
 correlator which, at this order, corresponds to the non-equilibrium 
 in-medium photon polarization tensor $\Pi_{\mu\nu}$.}
\end{figure}

We compute the relevant two-point function in Eq.\ (\ref{gnumber1}) 
using the method described in the previous section and treating
electromagnetic interactions in perturbation theory. At each order, 
there are two types of contributions: those which only involve 
the leptonic current $j^\mu$ and those which involve the emitting 
current $J^\mu$ at least once, which we shall call ``medium'' 
contributions. We will show in section~\ref{sec:app} that the 
former vanish identically when the dressing of the initial state 
is taken into account. Therefore, we only consider medium contributions 
here. The first non-trivial contribution to the connected correlator 
(\ref{gprop}) occurs at ${\cal O}(e^2)$ 
and involve the current $J^\mu$ twice. It is graphically represented 
on Fig.\ \ref{fig:gamma1}. There, the wavy lines represent the free 
photon propagator, that is, in Fourier space, the connected correlator 
of Fourier modes of photon field operators:
\beq
\label{freegprop}
 G^{\mu\nu} (t,t';\bk) = 
 \Big< \T_\C \Big\{ A_\emg^\mu (t,\bk)\,
 A_\emg^{\nu\,\dagger} (t',\bk) \Big\} \Big>_\emg\, ,
\eeq
where we have renamed the interaction picture photon field operators 
(see Eq.\ (\ref{intevol})) such as to emphasize that the corresponding 
time evolution is driven by the free electromagnetic Hamiltonian alone:
\beq
 A^\mu_\emg (t,\bk) \equiv A^\mu_\I (t,\bk)
 =\e^{iH_\emg t} \, A^\mu(\bk) \, \e^{-iH_\emg t} \, .
\eeq 
The subscript `em' on the bracket in Eq.\ (\ref{freegprop}) indicates
an averaging over electromagnetic degrees of freedom: 
$\langle \cdots \rangle_\emg \equiv \Tr[\rho_\emg \cdots]$. 
The blob of Fig.\ \ref{fig:gamma1} represents the out-of-equilibrium in-medium 
photon polarization tensor:
\beq
\label{polar}
 \Pi^{\mu\nu} (t,t';\bk) = 
 \Big< \T_\C \Big\{ J_\s^\mu (t,\bk)\,
 J_\s^{\nu\,\dagger}(t',\bk) \Big\} \Big>_\s\, ,
\eeq
where the subscript `s' on the bracket indicates that the average value 
is to be computed with the density matrix $\rho_\s$ of the system. Here,
$J_\s^\mu (t,\bk)$ is the interaction picture of the Fourier transform of 
the current operator $J^\mu (\bx)$ (see Eq.\ (\ref{intevol})):
\beq
\label{Jevol}
 J_\s^\mu (t,\bk)\equiv J_\I^\mu (t,\bk)
 =\e^{iH_\s t}\,J^\mu (\bk)\,\e^{-iH_\s t}\, .
\eeq
As in the case of the photon field above, we have renamed the interaction 
picture for this operator, which only depends on the degrees of freedom 
of the emitting system, to emphasize that its time dependence is 
completely determined by the Hamiltonian $H_\s$ of the system.
We stress, in particular, that, as far as the internal dynamics of 
the emitting system is concerned, $\Pi_{\mu\nu} (t,t';\bk)$ represents 
the {\em exact} unequal-time current-current correlator. 

The lowest order in-medium contribution to the photon propagator 
(\ref{gammaprop}) represented on Fig.\ \ref{fig:gamma1} reads:\footnote{The 
free part of the photon propagator does not contribute to the number
of produced photons and simply gives a constant contribution
to Eq.\ (\ref{gnumber1}), which is zero in the present case (see Eq.\ 
(\ref{gammavac})). Therefore, only the ``in-medium'' part to the photon 
propagator, Eq.\ (\ref{gammaprop}) below, contributes to Eq.\ (\ref{gnumber1}).}
\beq
\label{gammaprop}
 \G_{\mu\nu} (t,t';\bk) = -\int_\C  du dv\, G_{\mu\rho} (t,u;\bk)\,
 \Pi^{\rho\sigma} (u,v;\bk)\,G_{\sigma\nu} (v,t';\bk)\, ,
\eeq
with time integrals along the contour of Fig.\ \ref{fig:Kpath}.
The free vacuum photon propagator reads, in Feynman gauge:
\beq
\label{Fgauge}
 G_{\mu\nu} (t,t';\bq) = -g_{\mu\nu}\, G (t,t';\bk) \,,
\eeq
with (see Eqs.\ (\ref{correlsup}) and (\ref{correlinf}))
\bea
\label{freegpropsup}
 G^>(t,t';\bk)&=&\frac{1}{2k}\,\e^{-ik(t-t')}\\
\label{freegpropinf}
 G^<(t,t';\bk)&=&\frac{1}{2k}\,\e^{ik(t-t')} \, .
\eea
Writing explicitly the time integrals along the contour, we obtain:

\bea
\label{gammaprop2}
 \G_{\mu\nu}^< (t,t';\bk)= 
 -\int_{-\infty}^{t} [du] \int_{-\infty}^{t'} [dv] \,
 G_\rho(t,u;\bk) \, \Pi^<_{\mu\nu} (u,v;\bk) \, G_\rho (v,t';\bk)&\nonumber\\
 +\int_{-\infty}^{t} [du] \int_{-\infty}^{u} [dv] \,
 G_\rho(t,u;\bk) \, \Pi^\rho_{\mu\nu} (u,v;\bk) \, G^< (v,t';\bk)&\nonumber\\
 +\int_{-\infty}^{t'} [du] \int_{u}^{t'} \,\,[dv] \,
 G^<(t,u;\bk) \, \Pi^\rho_{\mu\nu} (u,v;\bk) \, G_\rho (v,t';\bk)
 &\!\!\!\!\!\!,
\eea
where we used the notation (\ref{timeint}) and where we introduced the 
so-called spectral component of the various two-point functions:
\beq
\label{spectralpi}
 \Pi_{\mu\nu}^\rho (t,t';\bk)=
 i\,\Big(\Pi_{\mu\nu}^> (t,t';\bk)-\Pi_{\mu\nu}^< (t,t';\bk)\Big)
\eeq
and similarly for the free propagator:
\beq
\label{freegproprho}
 G_\rho (t,t';\bk)=i\,\Big(G^> (t,t';\bk)-G^< (t,t';\bk)\Big)
 =\frac{\sin k(t-t')}{k}\, .
\eeq
Working out the time derivatives in Eq.\ (\ref{gnumber1}), we obtain 
the following expression for the single-photon inclusive spectrum at 
time $t$:
\beq
\label{gnumber}
 2k\,\frac{dn_\gamma (t,\bk)}{d^3x\,d^3k} = \frac{1}{(2\pi)^3}\,
 \int_{-\infty}^t [du][dv]\,\e^{-ik(u-v)}\,
 \gamma^{\mu\nu} (\bk)\,\Pi^<_{\nu\mu} (v,u;\bk)\, .
\eeq

We end this subsection by mentioning that the above expression 
(\ref{gammaprop2}) for the in-medium contribution to the photon 
propagator can be rewritten in a more compact form, which will prove 
useful for later use:
\bea
\label{gammaprop3}
 \G_{\mu\nu}^< (t,t';\bk) =
 \int_{-\infty}^{+\infty} [du][dv]\,
 \Big\{\,\,G_R(t,u;\bk)\,\Pi^<_{\mu\nu} (u,v;\bk) \, G_A (v,t';\bk)&\nonumber\\
 +G_R(t,u;\bk)\,\Pi^R_{\mu\nu} (u,v;\bk) \, G^< (v,t';\bk)&\nonumber\\
 +G^<(t,u;\bk)\,\Pi^A_{\mu\nu} (u,v;\bk) \, G_A (v,t';\bk)&\!\Big\}\,,
\eea
where we introduced the retarded and advanced components:
\bea
\label{retarded}
 \Pi^R_{\mu\nu} (t,t';\bk)&=&\Theta (t-t')\,\Pi^\rho_{\mu\nu} (t,t';\bk)\\
\label{advanced}
 \Pi^A_{\mu\nu} (t,t';\bk)&=&-\Theta (t'-t)\,\Pi^\rho_{\mu\nu} (t,t';\bk)\,,
\eea
and similarly for the free propagator $G$.

\subsection{Alternative derivation and physical interpretation}
\label{alternative}

Here, we present an alternative derivation of the 
formula (\ref{gnumber}) for the inclusive photon spectrum, not using 
the virtual negative time evolution formalism introduced in the 
previous section. For this purpose, we work in the basis 
where the density matrix $\rho_\s$ of the system is diagonal
(we use Greek letters to label the corresponding states
throughout this subsection): 
\beq
 \rho_\s = \sum_\alpha P^\s_\alpha |\alpha,\s\rangle \langle\alpha,\s|\, ,
\eeq
where $\langle \alpha,\s|\,\rho_\s\,|\beta,\s\rangle = 
\delta_{\alpha\beta}\,P^\s_\alpha$.

Let us first focus on a particular state $|\alpha,\s\rangle$ of this 
statistical ensemble and construct the corresponding dressed 
initial state $|\alpha\rangle$ at time $t=0$. At $\Op(e)$, 
the latter consists of a superposition of the bare zero-photon 
state and of all one-photon states allowed by the dynamics (the photon 
cloud). Making use of the adiabatic representation (\ref{adiab}), 
it can be expressed as:
\bea
\label{initstate}
 |\alpha\rangle
 &=&\Big\{1-i\int_{-\infty}^0 du\,\e^{\epsilon u}\,H_\I (u)\Big\}
 |\alpha,\s\rangle\otimes|0,\emg\rangle\nonumber\\
 &=&|\alpha,\s\rangle\otimes|0,\emg\rangle
 +V\!\int \frac{d^3k}{(2\pi)^3}\sum_{\lambda=0}^3 \sum_\beta\,
 \Phi^\pol_{\beta\alpha} (\bk)\,
 |\beta,\s\rangle\otimes|(\lambda,\bk),\emg\rangle
\eea
where
\beq
 |(\lambda,\bk),\emg\rangle\equiv a_\lambda^\dagger(\bk) |0,\emg\rangle
\eeq 
is the one-photon state with polarization $\lambda$ and momentum $\bk$.
The wave-function $\Phi^\pol_{\alpha\beta} (\bk)$ is given 
by the appropriate matrix element of the interaction Hamiltonian
(\ref{JA}), which reads explicitly: 
\beq
\label{wavefunc}
 i\Phi^\pol_{\beta\alpha} (\bk)=\int_{-\infty}^0 du\,\e^{\epsilon u}\,
 \e^{iku}\frac{\epsilon^\pol_\mu (\bk)}{\sqrt{2k}}\,
 \langle \beta,\s|J_\s^\mu (u,\bk)|\alpha,\s\rangle\, .
\eeq
The probability amplitude of having at least one photon 
with (transverse) polarization $\lambda$ and momentum $\bk$ at time $t$ is 
obtained by evolving the state (\ref{initstate}) up to time $t$ with the full 
Hamiltonian $H$ and by projecting over all possible final states 
containing at least the desired photon. At leading order in $e$,
only one-photon final states contribute and the associated probability 
amplitude consists of two distinct pieces corresponding to 
the cases where: (1) the measured photon is actually produced during 
the time evolution from $t=0$ to $t$ ; (2) the measured photon was 
already present in the initial virtual cloud. Below, we construct 
the two corresponding amplitudes at lowest order in $e$:

(1) {\it the measured photon is produced between times $t=0$ and $t$.} 
This corresponds to the case where one starts in the zero-photon part 
$|\alpha,\s\rangle\otimes |0,\emg\rangle$ of the initial state 
(\ref{initstate}). The corresponding amplitude can be obtained by the 
following sequence: First, the unperturbed time evolution from the 
initial state $|\alpha,\s\rangle\otimes |0,\emg\rangle$ at $t=0$ to 
an intermediate state $|i,\s\rangle\otimes |0,\emg\rangle$ at the 
intermediate time $0\le u\le t$. The associated amplitude is given 
by the corresponding matrix element of the unperturbed evolution
operator $\exp(-i[H_\s+H_\emg]t)$, which reduces to:
\beq 
 \langle i,\s| \e^{-iH_\s \,u} |\alpha,\s\rangle \,.
\eeq
Second, the transition from the state $|i,\s\rangle\otimes |0,\emg\rangle$ 
to one of the possible one-photon states 
$|j,\s\rangle\otimes |(\lambda,\bk),\emg\rangle$ allowed by the dynamics.
The corresponding transition amplitude is given by the matrix element of 
the interaction Hamiltonian (\ref{JA}) between these states: 
\beq
 \frac{\epsilon^\pol_\mu (\bk)}{\sqrt{2k}}\,
 \langle j,\s|J^\mu (\bk)|i,\s\rangle\, .
\eeq
Finally, the time-evolution 
from state $|j,\s\rangle\otimes |(\lambda,\bk),\emg\rangle$ at time $u$ to 
the final state $|\beta,\s\rangle\otimes |(\lambda,\bk),\emg\rangle$,
at time $t$, with the amplitude
\beq 
 \e^{-ik(t-u)} \,\langle \beta,\s| \e^{-iH_\s (t-u)} |j,\s\rangle \, .
\eeq
where the phase factor corresponds to the free propagation of the 
one-photon state.

Integrating over all possible times of emission and summing over all 
possible intermediate states, one obtains the amplitude:
\beq
 {\cal A}^\pol_{1,\alpha\rightarrow\beta} (t,\bk)=
 \frac{1}{\sqrt{2k}}\,\int_0^t du \, \e^{-ik(t-u)}\,
 \langle \beta,\s|\e^{-iH_\s t}\,\epsilon^\pol (\bk)\cdot J_\s (u,\bk)|\alpha,\s\rangle\, .
\eeq
where we used the definition (\ref{Jevol}).

(2) {\it the measured photon is part of the initial virtual cloud.} 
In that case, it freely propagates from time $t=0$ to time $t$.
The amplitude corresponding to the transition from a given
one-photon states $|\delta,\s\rangle\otimes|(\lambda,\bk),\emg\rangle$
of the virtual cloud in (\ref{initstate}) at time $t=0$ to the final
state $|\beta,\s\rangle\otimes |(\lambda,\bk),\emg\rangle$ at time $t$
reads:
\beq 
 \e^{-ikt} \,\langle \beta,\s| \e^{-iH_\s t} |\delta,\s\rangle \, .
\eeq
Summing over the possible states of the virtual cloud weighted with 
the associated wave-function (\ref{wavefunc}), one obtains the 
final amplitude:
\bea
 {\cal A}^\pol_{2,\alpha\rightarrow\beta} (t,\bk)&=&
 \sum_\delta\,\e^{-ikt}\,\langle \beta,\s| \e^{-iH_\s t} |\delta,\s\rangle
 \times \Phi^\pol_{\delta\alpha} (\bk)\nonumber\\
 &=&\frac{1}{\sqrt{2k}}\,\int_{-\infty}^0 du\,\e^{\epsilon u}\,\e^{-ik(t-u)}\,
 \langle \beta,\s|\e^{-iH_\s t}\,\epsilon^\pol (\bk)\cdot J_\s (u,\bk)|\alpha,\s\rangle\, .
\eea
Putting everything together we get, for the total amplitude:
\bea
 {\cal A}^\pol_{\alpha\rightarrow\beta} (t,\bk)&=&
 {\cal A}^\pol_{1,\alpha\rightarrow\beta} (t,\bk)
 +{\cal A}^\pol_{2,\alpha\rightarrow\beta} (t,\bk)\nonumber\\
 &=&\frac{1}{\sqrt{2k}}\, \int_{-\infty}^t [du] \, \e^{-ik(t-u)}\,
 \langle \beta,\s|\e^{-iH_\s t}\,\epsilon^\pol (\bk)\cdot J_\s (u,\bk)|\alpha,\s\rangle\, .
\eea
Squaring this amplitude to get the associated probability and summing 
over all possible final states $|\beta,\s\rangle$ as well as over transverse 
polarizations, one obtains the desired inclusive spectrum corresponding 
to the initial state (\ref{initstate}). Finally, averaging over all 
possible initial states $|\alpha,\s\rangle$ with the appropriate weight 
$P^\s_\alpha$, we recover our previous result,
Eq.\ (\ref{gnumber}), for the inclusive spectrum:
\beq
\label{gproba}
 (2\pi)^3\,\frac{dn_\gamma (t,\bk)}{d^3x\,d^3k}=
 \sum_{\alpha}P^\s_\alpha\,\left(\sum_{\lambda=\perp}
 \sum_\beta \Big| {\cal A}^\pol_{\alpha\rightarrow\beta} 
 (t,\bk)\Big|^2\right) 
\eeq

Therefore, we see that the various contributions to Eq.\ (\ref{gnumber})
corresponding to the various parts of the time integrations can be given
very simple interpretations. Writing schematically (we omit the
$\epsilon$--terms):
\beq
 \int_{-\infty}^t [du]\,[dv]\equiv
 \int_{-\infty}^0 du\,\int_{-\infty}^0 dv\,
 +\int_{-\infty}^0 du\,\int_0^t dv
 +\int_0^t du\int_{-\infty}^0 dv\,
 +\int_0^t du\int_0^t dv\,,
\eeq
we see, in the light of the previous discussion, that the contribution 
from negative times ($u,v\le 0$) corresponds to the probability that the 
photon in the final state was already present in the initial virtual 
cloud. Similarly, the contribution from purely positive times ($u,v\ge 0$)
corresponds to the probability that the measured photon has actually 
been produced between times $t=0$ and $t$. Finally, the cross-terms, 
which involve integrations over both negative and positive times, 
describe the interference between these two possibilities. Notice that 
the latter depends explicitly on time and, as a consequence, 
contributes to the actual production rate (see below).

\subsection{Photon production rate and spectrum of produced photons}
\label{sec:photonrate}
One obtains the out-of-equilibrium photon production rate at a given time 
$t$, by taking the time-derivative of Eq.\ (\ref{gnumber}). The complete 
expression at lowest order in $e$ is, therefore:
\beq
\label{grate}
 2k\,\frac{dn_\gamma (t,\bk)}{d^4x\,d^3k} = \frac{1}{(2\pi)^3}\,
 \int_{-\infty}^t [du]\,2\Re\Big\{\e^{ik(t-u)}\,
 \gamma^{\mu\nu} (\bk)\,\Pi^<_{\mu\nu} (t,u;\bk)\Big\} \, ,
\eeq
where we used the fact that: $\Pi^<_{\mu\nu} (t,t';\bk)=
\Pi^{<\,*}_{\nu\mu} (t',t;\bk)$. Equation (\ref{grate})
relates the photon production rate to the intrinsic
dynamical properties of the out-of-equilibrium emitting system,
which are characterized at this order by the unequal-time photon 
polarization tensor $\Pi^<_{\mu\nu} (t,u;\bk)$. This generalizes 
the corresponding equilibrium 
formula~\cite{Feinberg:1976ua,Gale:1990pn,McLerran:1984ay} 
(see Eq.\ (\ref{ttigrate}) below). 

The inclusive spectrum of actually {\em produced} photons can be obtained 
from Eq.\ (\ref{gnumber}) by subtracting the contribution from the initial 
virtual cloud, which is simply given by Eq.\ (\ref{gnumber}) itself, 
written at $t=0$:\footnote{Strictly speaking, the dressed initial state 
constructed in section \ref{sec:noneq} contains real radiation, which
should also be subtracted in order to obtain the amount of produced radiation. 
However, the number of real photons in the initial state can be made negligible 
by choosing the parameter $\epsilon$ such that: $\epsilon/\bar{k}\gg e^2$, where 
$\bar{k}$ is a typical scale in the problem. 
As mentioned previously, the present perturbative analysis at leading-order 
is not affected by this choice as long as $\epsilon/\bar{k}\ll e$.
[As an illustration of the previous point, consider a time-translation 
invariant situation, as discussed in section \ref{sec:app} below. In that
case, the number of real photon in the initial state can be shown to be 
$\sim {\mbox{$\bar \Pi$}^{<\,\mu}}_\mu (\omega=k,\bk)/k\epsilon$, where 
$\bPi^<_{\mu\nu}(\omega,\bk)$ is the photon polarization tensor in frequency 
space (see Eq.\ (\ref{frequencyFT})). One typically has
$\bar{\Pi}^{<\,\mu}_\mu(\omega=k,\bk)=e^2\,\bar{k}^2\,f(k/\bar{k})$, where 
the function $f(x)$ vanishes rapidly at large $x$. For instance, for a
system in equilibrium at temperature $T\equiv\bar{k}$, one has $f(x)\sim\exp(-x)$. 
Therefore, the contribution from real radiation in the initial state can be
neglected for $k\gtrsim\bar{k}$ by keeping $\epsilon/\bar{k}\gg e^2$.]} 
\beq
 \frac{dn_\gamma (t,\bk)}{d^3x\,d^3k}\Big|_{\rm prod} \equiv 
 \frac{dn_\gamma (t,\bk)}{d^3x\,d^3k}-\frac{dn_\gamma (0,\bk)}{d^3x\,d^3k}
\eeq
This is, of course, equivalent to a direct integration of the 
production rate (\ref{grate}) from $t=0$ to $t$, namely:
\beq
 \frac{dn_\gamma (t,\bk)}{d^3x\,d^3k}\Big|_{\rm prod}
 =\int_0^t dt'\, \frac{dn_\gamma (t',\bk)}{d^4x\,d^3k}\,.
\eeq

Notice that, although the contribution from the virtual cloud has been 
subtracted, there still remains a time integral over infinite negative 
times in Eq.\ (\ref{grate}). As we have seen previously, this 
corresponds to interference effects with the virtual cloud.

\subsection{Inclusive spectrum of correlated lepton pairs at time $t$}

\begin{figure}[t]
 \begin{center}
 \epsfig{file=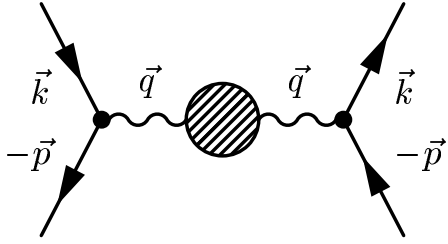,width=6.cm}
 \end{center}
 \caption{\label{fig:dilep1}
 The lowest order medium contribution to dilepton production.
 The dashed blob represents the ${\cal O}(e^2)$ in-medium photon 
 polarization tensor. Lepton momenta follow the fermionic flow.}
\end{figure}

We now come to the case of dilepton production. The density of {\em correlated}
lepton pairs with momenta $\bk$ (lepton) and $\bp$ (anti-lepton) at time $t$ 
is given by 
\beq
\label{llspectrum}
 \frac{dn_\dl (t,\bk,\bp)}{d^3x\,d^3k\,d^3p} =
 \frac{V}{(2\pi)^6}\, \sum_{\sigma,\sigma'} 
 \Big< N^{\ell^-}_\sigma (t,\bk)\,N^{\ell^+}_{\sigma'} (t,\bp)\Big> \, ,
\eeq
where the RHS is to be understood as a {\em connected} correlator.
Here $V\equiv (2\pi)^3\,\delta^{(3)} (\vec 0)$ is the quantization 
volume and 
\bea
  N^{\ell^-}_\sigma (t,\bp) &=& b_\sigma^\dagger (t,\bp)\, b_\sigma (t,\bp)\\
  N^{\ell^+}_\sigma (t,\bp) &=& d_\sigma^\dagger (t,\bp)\, d_\sigma (t,\bp)
\eea
denote the Heisenberg picture number operators for 
leptons and anti-leptons with momentum $\bp$ and spin projection $\sigma=\pm$. 
In terms of the Fourier modes of the lepton field operators
\bea
  \psi (t,\bk) &=& \frac{1}{\sqrt V}\,
  \int d^3x\,\e^{-i\bk \cdot \bx} \, \psi (t,\bx)\\
  \bar{\psi} (t,\bk) &=& \frac{1}{\sqrt V}\,
  \int d^3x\,\e^{i\bk \cdot \bx} \, \bar{\psi} (t,\bx)\,,
\eea
the relevant correlator is given by the 
following connected four-point function: 
\bea
 \sum_{\sigma,\sigma'} \Big< N^{\ell^-}_\sigma (t,\bk)\,
 N^{\ell^+}_{\sigma'} (t,\bp)\Big>
 &=&\Big< \tr\Big\{\bar{\psi} (t,\bk)\,
 \gamma_0\,\frac{K\slash+m}{2E_k}\,\gamma_0\,\psi (t,\bk)\Big\}\nonumber\\
 &&\times \tr\Big\{\gamma_0\,\frac{P\slash-m}{2E_p}\,\gamma_0\,
 \psi (t,-\bp)\,\bar{\psi} (t,-\bp)\Big\}\Big>\, ,
\eea
where $m$ is the lepton mass, $E_k=\sqrt{k^2+m^2}$, $E_p=\sqrt{p^2+m^2}$,
$K^\mu\equiv(E_k,\bk)$, $P^\mu\equiv(E_p,\bp)$, $K\slash = \gamma^\mu K_\mu$ 
and $P\slash = \gamma^\mu P_\mu$.
This correlator can be computed using the method described in 
section~\ref{sec:noneq}. 
As in the case of photons the purely leptonic contributions, 
involving the lepton current $j^\mu$ only, vanish identically. 
Here, we focus on medium contributions, which involve $J^\mu$ at 
least once. The first non-vanishing medium contributions 
occur at ${\cal O}(e^4)$ and involve the unequal-time 
current-current correlator (\ref{polar}).
There are two possible topologies at this order, one of which is
represented on Fig.\ \ref{fig:dilep1}, where the lines represent 
the free lepton propagator on the contour $\C$:
\beq
 D(t,t';\bk)=\Big<T_\C\Big\{\psi_\emg (t,\bk) \,
 \bar{\psi}_\emg (t',\bk) \Big\}\Big>_\emg\,.
\eeq
As we did previously for the photon field, we have renamed the 
interaction picture of lepton field operators as follows:
\beq
 \psi_\emg (t,\bk)\equiv\psi_{\,\I} (t,\bk)
 =\e^{iH_\emg t}\,\psi (\bk)\,\e^{-iH_\emg t}\, ,
\eeq
and likewise for $\bar{\psi}_\emg (t,\bk)$. With initial
conditions corresponding to Eq.\ (\ref{gammavac}), the components
of the lepton propagator are given by:
\bea
\label{bfermion1}
 D^>(t,t';\bk)&=&\frac{K\slash+m}{2E_k}\,\e^{-iE_k(t-t')}\\
\label{bfermion2}
 D^<(t,t';-\bk)&=&\frac{K\slash-m}{2E_k}\,\e^{iE_k(t-t')} \, .
\eea

The contribution depicted in Fig.\ \ref{fig:dilep1} 
can be expressed in terms of the lowest order in-medium contribution to 
the photon propagator $\G_{\mu\nu}$, represented on Fig.\ \ref{fig:gamma1} 
(see Eq.\ (\ref{gammaprop})):
\beq
\label{ltensor}
 V\cdot\sum_{\sigma,\sigma'} \Big< N^{\ell^-}_\sigma (t,\bk)\,
 N^{\ell^+}_{\sigma'} (t,\bp)\Big>
 =-e^2\,\int_\C du dv\, l^{\mu\nu} (t,u,v;\bk,\bp)\,
 \G_{\nu\mu} (v,u;\bq)\,,
\eeq
where $\bq=\bk+\bp$ is the total momentum of the lepton pair and where
\bea
 l^{\mu\nu} (t,u,v;\bk,\bp)&=&\tr\Big\{D(t,u;-\bp)\gamma^\mu\,D(u,t;\bk)\,
 \gamma_0\,\frac{K\slash+m}{2E_k}\,\gamma_0\nonumber\\
 &&\quad\times D(t,v;\bk)\,\gamma^\nu\,D(v,t;-\bp)\,
 \gamma_0\,\frac{P\slash-m}{2E_p}\,\gamma_0\,\Big\} \,.
\eea
Writing explicitly the time integrations along the contour and making
use of Eqs.\ (\ref{bfermion1})-(\ref{bfermion2}), one can perform the 
trace over Dirac indices. A lengthy but straightforward calculation 
shows that all but one of the contributions on the contour vanish. 
A similar calculation shows that the other possible contribution to 
the fermion four-point function gives a vanishing contribution to 
dilepton production after performing the relevant projections on 
positive and negative energy final states. 
Putting everything together, we finally obtain the following expression 
for the inclusive spectrum of correlated lepton pairs at time $t$, at 
leading order in the electromagnetic coupling constant:
\beq
\label{llnumber1}
 \frac{dn_\dl (t,\bk,\bp)}{d^3x\,d^3k\,d^3p} =
 \frac{e^2}{(2\pi)^6}\,\int_{-\infty}^t [du][dv] \,
 \e^{-i(E_k+E_p)(u-v)}\,\ell^{\mu\nu} (\bk,\bp)\,\G^<_{\nu\mu} (v,u;\bq)\,,
\eeq
where
\beq
 \ell_{\mu\nu} (\bk,\bp)
 =\frac{K_\mu P_\nu+P_\mu K_\nu - g_{\mu\nu}(K\cdot P+m^2)}{E_k\,E_p}\, .
\eeq
is the usual vacuum lepton tensor~\cite{LeBellac:book}. One
obtains the number of correlated pairs with total four-momentum 
$Q^\mu=(q_0,\bq)$ by integrating (\ref{llnumber1}) over momenta 
with the constraints $\bp+\bk=\bq$ and $E_p+E_k=q_0$. Using
\beq
 \int \frac{d^3k}{(2\pi)^3}\frac{d^3p}{(2\pi)^3}\,
 (2\pi)^4\,\delta^{(4)} (Q-P-K)\,\ell_{\mu\nu} (\bk,\bp)\,
 \e^{i(E_k+E_p)t}=L_{\mu\nu} (Q^2)\,\e^{iq_0t}\, ,
\eeq
where ($Q^2\ge 4 m^2$)
\beq
\label{leptensor}
 L_{\mu\nu} (Q^2)=\frac{1}{6\pi}\,\Big(1+\frac{2m^2}{Q^2}\Big)\,
 \Big(1-\frac{4m^2}{Q^2}\Big)^{\!\frac{1}{2}}\,[Q_\mu Q_\nu-g_{\mu\nu}Q^2]\, ,
\eeq
we obtain:
\beq
\label{llnumber2}
 \frac{dn_\dl (t,q_0,\bq)}{d^3x\,d^4Q} =
 \frac{e^2}{(2\pi)^4}\,\int_{-\infty}^t [du][dv] \,
 \e^{-iq_0(u-v)}\,L^{\mu\nu} (Q^2)\,\G^<_{\nu\mu} (v,u;\bq)\, .
\eeq

Note that the general structure of the expressions (\ref{llnumber1}) and 
(\ref{llnumber2}) is very similar to that obtained in the case of photon 
production, Eq.\ (\ref{gnumber}). In particular, the integrand does not 
explicitly depend on $t$ and the time-dependence only appears through 
the upper bound of the time integrations. Following the line of reasoning 
of section~\ref{alternative}, the various integrations over negative and 
positive times involved in Eqs.\ (\ref{llnumber1}), (\ref{llnumber2}) and 
(\ref{gammaprop2}) can be given simple physical interpretations. 
At lowest order in $e$, the basic process is the emission of an off-shell 
photon which decays into a lepton pair and the probability amplitude 
for having (at least) a lepton pair in the final state can be decomposed 
in various pieces corresponding to the various possible times of emission
and decay of the off-shell photon. For instance, the pair may already be 
present in the virtual cloud, which correspond to both times being 
negative, or an off-shell photon from the virtual cloud may decay 
between $t=0$ and $t$, etc. The various pieces of the time integrations
in Eqs.\ (\ref{llnumber1}), (\ref{llnumber2}) and (\ref{gammaprop2}) correspond 
either to the probabilities associated with each possibility or to the 
interferences between different possibilities.

\subsection{Dilepton production rate and spectrum of produced pairs}

As in the case of photon production, one obtains the rate of lepton pair
production as a function of time by taking the time derivative of expressions 
(\ref{llnumber1}) and (\ref{llnumber2}) above. We get, for pairs of lepton 
and anti-lepton with momenta $\bk$ and $\bp$ respectively:
\beq
\label{llrate1}
 \frac{dn_\dl (t,\bk,\bp)}{d^4x\,d^3k\,d^3p} =
 \frac{e^2}{(2\pi)^6}\,\int_{-\infty}^t [du] \,
 2\Re\Big\{\e^{i(E_p+E_k)(t-u)}\,\ell^{\mu\nu} (\bk,\bp)\,
 \G^<_{\mu\nu} (t,u;\bq)\Big\}\, ,
\eeq
and, for a pair with total four momentum $Q$:
\beq
\label{llrate2}
 \frac{dn_\dl (t,q_0,\bq)}{d^4x\,d^4Q} =
 \frac{e^2}{(2\pi)^4}\,\int_{-\infty}^t [du] \,
 2\Re\Big\{\e^{iq_0(t-u)}\,L^{\mu\nu} (Q^2)\,\G^<_{\mu\nu} (t,u;\bq)\Big\}\, .
\eeq
The number of produced pairs at time $t$ is obtained either by subtracting
the virtual contribution ($t=0$) from expressions (\ref{llnumber1}) and 
(\ref{llnumber2}) or by integrating the above rates from $t=0$ to $t$. For
instance:
\beq
 \frac{dn_\dl (t,q_0,\bq)}{d^3x\,d^4Q}\Big|_{\rm prod} = 
 \int_0^t dt' \,\frac{dn_\dl (t',q_0,\bq)}{d^4x\,d^4Q}\, .
\eeq

\subsection{Charged scalar fields}

We end this section by considering how the above formulas for the photon
and dilepton production rates are modified in the case where the
emitting system contains charged scalar fields such as e.g.\ charged pions.
This is of interest e.g.\ in the hadronic phase in heavy ion collisions, or 
during the out-of-equilibrium chiral phase transition, where so-called 
disoriented chiral condensates might occasionally form, leading to 
coherent pion emission~\cite{Rajagopal:1993ah}. The accompanied 
electromagnetic radiation might provide an interesting signature of 
this phenomenon~\cite{Huang:1996kq,Boyanovsky:1997cy}.

Apart from the contribution of charged scalar field to the current
(\ref{emcurrent}), the electromagnetic interaction Hamiltonian (\ref{JA}) 
now contains an additional contribution of the type:  
\beq
 H_{\rm int}^{\rm scal}= e^2\,\int d^3x\, A^\mu(\bx)A_\mu(\bx)\,
 \phi^\dagger (\bx) \phi (\bx)\, ,
\eeq
where $\phi (\bx)$ denotes the charged scalar field operator. 
Repeating the above analysis for photon and dilepton production, it 
is easy to see that, at lowest order in $e$, the presence of this 
``two-photon'' vertex simply amounts to an additional tadpole-like 
contribution to the expression (\ref{gammaprop}) of the in-medium 
photon polarization tensor. It can be formally obtained from 
Eq.\ (\ref{gammaprop}) by replacing $\Pi_{\mu\nu} (u,v;\bq)$ by 
$-i\,{\rm K} (u)\,g_{\mu\nu}\,\delta_\C(u-v)$, where 
\beq
 {\rm K} (t)=e^2 \,\int \frac{d^3 k}{(2\pi)^3}\,
 \langle \varphi_\s^\dagger(t,\bk)\,\varphi_\s (t,\bk)\rangle_\s\, ,
\eeq
with $\varphi_\s (t,\bk)$ the interaction picture of the Fourier modes 
of the field operator (see Eq.\ (\ref{int1})):
\beq
 \varphi_\s (t,\bk)\equiv\varphi_\I (t,\bk)
 =\e^{iH_\s t}\,\varphi (\bk)\,\e^{-iH_\s t}\, ,
\eeq
where ($V$ is the normalization volume)
\beq
 \varphi (\bk)=\frac{1}{\sqrt V}\,
 \int d^3x \, \e^{-i\bk\cdot\bx}\,\phi (\bx)\, .
\eeq
Writing explicitly the time integrals along the contour, one obtains that, 
in presence of scalar fields, the tadpole contribution to Eq.\ 
(\ref{gammaprop2}) reads:
\bea
\label{tadpole}
 \G_{\mu\nu}^< (t,t';\bq)\Big|_{\rm tadpole} 
 &=&g_{\mu\nu}\,\int_{-\infty}^{t} [du] \,G_\rho(t,u;\bq)\,
 {\rm K}(u)\,G^< (u,t';\bq)\nonumber\\
 &-&g_{\mu\nu}\,\int_{-\infty}^{t'} [du]\,G^<(t,u;\bq)\,
 {\rm K}(u)\,G_\rho (u,t';\bq)\,.
\eea

Using the expressions (\ref{freegpropsup}), (\ref{freegpropinf}) and 
(\ref{freegproprho}) of free propagators, it is straightforward to show 
that the tadpole term does not contribute to photon production:
\beq
 \lim_{t'\rightarrow t}\,\,
 \Big(\partial_t\partial_{t'}+ik(\partial_t-\partial_{t'})+k^2\Big)\,
 \G_{\mu\nu}^< (t,t';\bk)\Big|_{\rm tadpole}=0\, .
\eeq
However, in general it gives a non-vanishing contribution to the 
out-of-equilibrium dilepton production. We stress that this 
contribution results from the non-trivial time-evolution of the 
emitting system. In particular, it vanishes identically in equilibrium 
and, more generally, for any stationary state, as we will show below.
It is, therefore, a genuine non-equilibrium effect. 

\section{Applications}
\label{sec:app}

In this section, we apply the general formulas derived above to the
case of time-translation invariant (stationary) systems. 
In particular, we recover known expressions for the production rates in 
thermodynamic equilibrium. We also show that a consistent
description of the initial virtual cloud guarantees that only
on-shell processes contribute to production rates in the case
of stationary systems and that the vacuum is stable against 
spontaneous emission. These provide important checks of the physical 
consistency of the present description.
Finally, we consider the case of quasi-stationary systems and work out,
in particular, the first non-trivial corrections to the local expression
of the photon production rate -- used e.g.\ in hydrodynamic calculations -- 
in a standard gradient expansion. 

\subsection{Recovering known formulas for stationary systems}

It is instructive to reproduce the usual equilibrium expressions
for the photon and dilepton production rates from the general 
out-of-equilibrium expressions derived above. For this purpose, we 
consider the case where the emitting system undergoes a stationary
evolution. This is the case whenever the initial condition does not 
break time-translation invariance, that is when the density matrix 
$\rho_\s$ commutes with the Hamiltonian $H_\s$ (it can, therefore, 
be diagonalized in the basis of eigenstates $|n,\s\rangle$, cf. 
Eq.\ (\ref{dmsyst})):
\beq
  [\rho_\s ; H_\s ] = 0 \,\Leftrightarrow\, 
  P^\s_{nm}=P^\s_n \, \delta_{m,n}\, .
\eeq
In that case, all correlation functions only depend on time differences 
and it is useful to work in frequency space.
In particular, one has:
\beq
\label{frequencyFT}
 \Pi^<_{\mu\nu} (t,t';\bk)\equiv\Pi^<_{\mu\nu} (t-t';\bk)
 = \int_{-\infty}^{+\infty} \frac{d\omega}{2\pi}\,
 \e^{-i\omega(t-t')}\,\bPi^<_{\mu\nu} (\omega,\bk)\, ,
\eeq
and similarly for other two-point functions. Note the relation 
\beq
\label{relation}
 \bPi^{<\,*}_{\mu\nu} (\omega,\bk)=\bPi^<_{\nu\mu} (\omega,\bk)\,.
\eeq
Plugging the definition (\ref{frequencyFT}) in Eq.\ (\ref{grate}) 
and using the identity:\footnote{Writing
$F (t)=\int_{-\infty}^t[du]\,\e^{i\alpha (t-u)}$, 
with the definition (\ref{timeint}), one finds that the time 
derivative $dF/dt=\lim_{\epsilon \rightarrow 0^+}\,
i\,\epsilon\,\e^{i\alpha t}/(\alpha+i\epsilon)=0$.
It follows that $F(t)=F(0)=i/(\alpha+i\epsilon)$.}
\beq
\label{negtime}
 \int_{-\infty}^t[du]\,\e^{i\alpha (t-u)}=
 \pi\,\delta (\alpha)+i\,{\cal P} \Big(\frac{1}{\alpha}\Big) \, ,
\eeq
where $\cal P$ denotes the principal part, one obtains the following 
expression for the (time independent) photon production rate:
\beq
\label{ttigrate}
 2k\,\frac{dn_\gamma (\bk)}{d^4x\,d^3k} = 
 \frac{1}{(2\pi)^3}\,\gamma^{\mu\nu} (\bk)\,\bPi^<_{\mu\nu} (k,\bk)=
 -\frac{1}{(2\pi)^3}\,{\mbox{$\bar \Pi$}^{<\,\mu}}_\mu (k,\bk)\, .
\eeq
In writing the second equality, we used the fact that $\bPi_{\mu\nu}$
is transverse ($K^\mu=(k^0,\bk)$):
\beq
\label{transverse}
 K^\mu\,\bPi_{\mu\nu}(k_0,\bk)=0\,,
\eeq
as a consequence of gauge invariance. 

Similar considerations directly lead to the following formula for the
production rate of dileptons with invariant mass $Q$ 
(see Eq.\ (\ref{llrate2})):
\beq
 \frac{dn_\dl (q_0,\bq)}{d^4x\,d^4Q} =
 \frac{e^2}{(2\pi)^4}\,L^{\mu\nu} (Q^2)\,\bcG^<_{\mu\nu} (q_0,\bq)\, ,
\eeq
where the function $\bcG^<$ is defined as in Eq.\ (\ref{frequencyFT}). It 
can be easily expressed in terms of the polarization tensor $\bPi^<$ by
using Eq.\ (\ref{gammaprop3}) which has the form of a convolution product
in time and which, therefore, corresponds to an algebraic product in frequency 
space. This is, of course, nothing else but the fact that time-translation
invariance is equivalent to energy (frequency) conservation. The various 
components of the free photon propagator (\ref{Fgauge}) have the following 
expressions in frequency space: 
\bea
\label{Fourierfree1}
 \bG^< (\omega,\bq) &=& \frac{\pi}{q}\,\delta(\omega+q)\\
 \bG^R (\omega,\bq) &=& \frac{-1}{(\omega+i\eta)^2-q^2}\\
\label{Fourierfree2}
 \bG^A (\omega,\bq) &=& \frac{-1}{(\omega-i\eta)^2-q^2}
\eea
where $\eta\equiv0^+$. The contributions corresponding to the last two terms 
of Eq.\ (\ref{gammaprop3}) contain $G^<_{\mu\nu} (q_0,\bq)\propto \delta (q_0+q)$
and, therefore, give vanishing contributions for $q_0>0$. The additional 
tadpole contribution appearing in the case of scalar fields (see 
Eq.\ (\ref{tadpole})) vanishes identically for the same reason.
Using Eqs.\ (\ref{leptensor}) and (\ref{transverse}), one finally 
obtains ($\alpha_\emg=e^2/4\pi$):
\beq
\label{ttillrate}
 \frac{dn_\dl (Q)}{d^4x\,d^4Q} =
 -\frac{\alpha_\emg}{24\pi^4\,Q^2}\,\Big(1+\frac{2m^2}{Q^2}\Big)\,
 \Big(1-\frac{4m^2}{Q^2}\Big)^{\!\frac{1}{2}}\,\,
 {\mbox{$\bar \Pi$}^{<\,\mu}}_\mu (q_0,\bq)\,.
\eeq
Equations.\ (\ref{ttigrate}) and (\ref{ttillrate}) are the usual expressions 
of the photon and dilepton production rates for general stationary 
situations. 
For the particular case of thermodynamic equilibrium 
at temperature $T=1/\beta$ these are usually expressed in terms 
of the spectral function 
$\bPi^\rho_{\mu\nu} (q_0,\bq)=2i\,\Im\bPi^R_{\mu\nu} (q_0,\bq)$ 
(see Eqs.\ (\ref{spectralpi}), (\ref{retarded}) and (\ref{advanced})) 
by making use of the detailed balance (or KMS) relation: 
\beq
 \bPi^>_{\mu\nu}(q_0,\bq)\Big|_{\rm eq}=
 \e^{\beta q_0}\,\bPi^<_{\mu\nu} (q_0,\bq)\Big|_{\rm eq}\,,
\eeq 
from which it follows that
\beq
\label{KMS}
 \bPi^<_{\mu\nu} (q_0,\bq)\Big|_{\rm eq} = 
 \frac{2}{\e^{\beta q_0}-1}\,\Im\bPi^R_{\mu\nu} (q_0,\bq)\Big|_{\rm eq} \, .
\eeq
Putting everything together, one recognizes the familiar expressions for 
the photon and dilepton production rates in 
equilibrium~\cite{Feinberg:1976ua,Gale:1990pn,McLerran:1984ay,LeBellac:book}. 

As an important physical consequence, the above results guarantee that
the vacuum is stable against spontaneous emission. Indeed, the fact that 
only on-shell (energy-conserving) processes contribute in stationary 
situations implies that:
\beq
\label{vacuum}
 {\mbox{$\bar \Pi$}^{<\,\mu}}_\mu (q_0,\bq)\Big|_{\rm vacuum} = 0\,,
\eeq
which can also be obtained as the zero temperature limit of 
Eq.\ (\ref{KMS}). As a byproduct, this has the important
implication that the purely leptonic contributions to the photon 
polarization tensor, which only involve the leptonic current $j^\mu$
and have, therefore, the structure of vacuum contributions, do
not contribute to actual photon and dilepton production rates,
as announced in section~\ref{sec:emrad}.

\subsection{Slowly evolving systems: gradient expansion and off-shell effects}

As mentioned in the introduction, a widely used ansatz
for computing photon or dilepton production rates out of equilibrium is
to use the above static formulas, Eqs\ (\ref{ttigrate}) or (\ref{ttillrate}), 
written at a given time $t$. One then supplement these local expressions for 
the production rates with a given modelization of the time-evolution of the 
system, like e.g.\ hydrodynamics, or an appropriate Boltzmann equation. 
This procedure might be justified if the typical time scale characterizing 
the process under study, say the formation time of the produced photon, 
is short compared to the time scale characterizing the time evolution of 
the system. In contrast, we clearly see from the expressions derived 
previously that in the general case the emission rates depend 
non-locally on the time history of the emitting system prior to the 
time of emission. It is therefore interesting to consider a somewhat 
intermediate situation where the system is slowly evolving (compared to 
the duration of the emission process) and to derive the first corrections 
to the static formulas, Eqs.\ (\ref{ttigrate}) and (\ref{ttillrate}). 
With the general expressions of the previous section in hand, we are 
in a position to work out these corrections by performing a standard 
gradient expansion. Here, we illustrate this point on the case of photon 
production.

Generalizing the notation introduced in Eq.\ (\ref{frequencyFT}), 
we introduce the Wigner transform in frequency space, that is
a Fourier transform with respect to the time difference $t-t'$
at fixed $T=\frac{1}{2}(t+t')$:
\beq
\label{WT}
 \Pi^<_{\mu\nu} (t,t';\bk)= \int \frac{d\omega}{2\pi}\,
 \e^{-i\omega (t-t')}\,\bPi^<_{\mu\nu} (T;\omega,\bk)\, ,
\eeq
and similarly for other two-point functions. Note that the relations
(\ref{relation}) and (\ref{transverse}) generalize:
\beq
 \bPi^{<\,*}_{\mu\nu} (T;\omega,\bk)=\bPi^<_{\nu\mu} (T;\omega,\bk)\, ,
\eeq
and
\beq
\label{transverse2}
 K^\mu\,\bPi_{\mu\nu} (T;k_0,\bk)=0\,.
\eeq
The photon production rate (\ref{grate}) can be rewritten as:
\beq
\label{grategrad}
 \frac{dn_\gamma (t,\bk)}{d^4x\,d^3k} = \int \frac{d\omega}{2\pi}\,
 \int_{-\infty}^t [du]\,\bF_\gamma (U;\omega,\bk) \, 
 2\Re\Big\{ \e^{i(k-\omega)(t-u)}\Big\},
\eeq
where $U=\frac{1}{2}(t+u)$ and where we introduced the (real) function
\beq
\label{gammafunc}
 \bF_\gamma(T;\omega,\bk) = 
 \frac{\gamma^{\mu\nu} (\bk)}{2k\,(2\pi)^3}\,
 \,\bPi^<_{\mu\nu} (T;\omega,\bk) \,.
\eeq
Writing $s=t-u$ and $U=t-\frac{1}{2}s$, and plugging the first order
gradient expansion
\beq
 \bF_\gamma (U;\omega,\bk)=
 \bF_\gamma (t;\omega,\bk) - \frac{1}{2}s\,\partial_t\bF_\gamma (t;\omega,\bk)
 +\ldots
\eeq
in Eq.\ (\ref{grategrad}), one obtains, after simple manipulations
and making use of Eq.\ (\ref{negtime}):
\beq
\label{gammagrad}
 \frac{dn_\gamma (t,\bk)}{d^4x\,d^3k} \simeq \bF_\gamma (t;k,\bk)
 +\PP\int\frac{d\omega}{2\pi}\,
 \frac{\partial_t\bF_\gamma (t;\omega,\bk)}{(k-\omega)^2} \,.
\eeq
Using Eqs.\ (\ref{gtensor}), (\ref{transverse2}) and (\ref{gammafunc}), 
the first term of this expression reduces to the static expression 
(\ref{ttigrate}) written at time $t$. This corresponds to the 
local ansatz described above. The second term represent the first non-trivial 
gradient correction arising from the finite duration of the emission 
process. As expected, it involves the off-shell part 
of the photon polarization tensor\footnote{Notice that one cannot make 
use of Eq.\ (\ref{transverse2}) under the frequency integral in Eq.\ 
(\ref{gammagrad}) and the full expression of the tensor $\gamma^{\mu\nu} (\bk)$ 
has to be kept.}, which in turn involves energy non-conserving elementary 
processes.

\section{The non-equilibrium photon polarization tensor}
\label{sec:polar}

In the previous sections, we have related the time-dependent photon 
and dilepton production rates to the intrinsic dynamical properties of the 
out-of-equilibrium emitting medium which, at leading order in $e$, are 
characterized by the unequal-time current-current correlator (\ref{polar}). 
In the last part of this paper, we focus on the latter and discuss how it 
might be computed in general non-equilibrium situations. We first consider 
the issue of the negative time integration which involves the unequal-time 
correlation function of two current operators at positive and negative times. 
We propose a general method for the calculation of such unusual correlation 
functions, which is based on introducing a $\Sp$--shape path in real time. 
This might be combined with recently developed techniques in non-equilibrium 
quantum field theory in order to compute the explicit time dependence of the 
relevant correlation functions starting from a given initial condition. 
Finally, we briefly discuss how the photon polarization tensor can be
obtained from the 2PI effective action.

\subsection{General considerations: the $\Sp$-path}

For the purpose of illustration, let us consider the case of photon 
production. Writing explicitly the various part of the time integration, 
the production rate (\ref{grate}) can be written as:
\beq
\label{gammanptime}
 \frac{dn_\gamma (t,\bk)}{d^4x\,d^3k} = 2\Re\left\{\,
 \int_0^t du\,\e^{ik(t-u)}\, {\cal F}_\gamma(t,u;\bk) +
 \int_{-\infty}^0 du\,\e^{\epsilon u}\,
 \e^{ik(t-u)}\,{\cal F}_\gamma(t,u;\bk)\,\right\} \, ,
\eeq
where we denoted ${\cal F}_\gamma(t,u;\bk)=
 \frac{\gamma^{\mu\nu}(\bk)}{2k(2\pi)^3}\,
 \Pi_{\mu\nu}(t,u;\bk)$.
As discussed previously, the first term on the RHS of Eq.\ 
(\ref{gammanptime}) describes the actual production of a photon
between time $t=0$ and time $t$, while the second term corresponds
to interferences with the initial virtual photon cloud. We see that
the latter involves the correlation $\Pi_{\mu\nu}(t,u;\bk)$ (see Eq.\ 
(\ref{polar})) between two current operators at positive time $t$ and 
negative time $u$ respectively. 
In practice, starting from a given initial condition, one therefore has 
to compute the time evolution of correlation functions of the emitting 
system in the positive as well as in the negative time direction (we 
recall that the latter is not a physical time evolution but merely a 
formal way of treating initial state correlations due to 
electromagnetic dressing). This can be done by introducing an 
appropriate contour in real time, such as, for instance, that of 
Fig.\ \ref{fig:Kpath}. The latter is, however, not well suited
for practical calculation as it requires one to specify 
the initial state at $t=-\infty$. Here we shall introduce a different contour
in real time which allows for a more useful formulation of the 
problem.\footnote{It is important to notice that the contour $\C$ 
introduced in section \ref{sec:noneq} only concerns insertions of 
electromagnetic vertices. We are free to choose any other contour to 
compute the internal dynamics of the emitting system.}

\begin{figure}[t]
 \begin{center}
 \epsfig{file=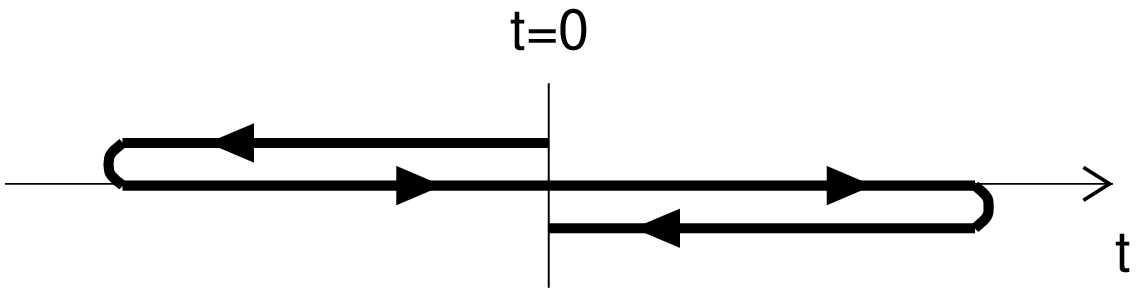,width=8.cm}
 \end{center}
 \caption{\label{fig:Spath}
 The $\Sp$--path in real time.}
\end{figure}

We consider a generic quantum field theory described by a set of fundamental 
(bosonic and/or fermionic) fields, which we collectively denote by $\varphi$.
For the sake of the argument, we focus on the time-evolution of the system
and omit all possible (Dirac, Lorentz, etc \ldots) indices as well as spatial 
variables. Let us consider a general correlation function of two operators 
$\Op$ at positive and negative times. We first notice that the operator with 
negative time argument can always be brought on the right of the one with 
positive time argument in the expressions of the production rates. The relevant 
correlation function can therefore be written ($t,t'\ge 0$):
\beq
 \langle \Op_\s (t)\,\Op_\s (-t')\rangle_\s=
 \sum_{n,m} P^\s_{nm} \,\langle m,\s|\,{\cal U} (0,t)\,\Op\,
 {\cal U} (t,0)\,{\cal U} (0,-t')\,\Op\,{\cal U} (-t',0)|n,\s\rangle\,,
\eeq\
with the time-evolution operator
\beq
 {\cal U} (t,t')=\e^{-iH_\s(t-t')}\,.
\eeq
The above expression can be read as follows: Starting from a given state
$|n,\s\rangle$ at time $t=0$, evolve the system backward in time up
to time $-t'$ and insert the operator $\Op$; Then, evolve the 
system in the positive time direction from time $-t'$ to time $t$
and insert $\Op$ again; Finally, evolve the system back from time
$t$ to time $t=0$ and compute the overlap with another state 
$|m,\s\rangle$. Repeat this for all possible states and average with 
the appropriate weight $P^\s_{nm}\equiv \langle n,\s|\,\rho_\s\,|m,\s\rangle$.
It is easy to show that the calculation of such correlation function can 
be formulated by means of standard field theory methods by introducing a 
closed $\Sp$--shape contour along the real time axis, such as represented 
on Fig.\ \ref{fig:Spath}.
In particular, this can be given a path-integral formulation along the
$\Sp$--path, allowing one to exploit the machinery of functional methods,
which have proven extremely powerful in recent years for the description 
of non-equilibrium systems. A general time-ordered $n$--point correlation
function with either positive or negative time-arguments on the contour 
$\Sp$ can be written as:
\beq
 \Big< T_\Sp \Big\{\Op_\s (t_1)\cdots\Op_\s (t_n)\Big\}\Big>_\s
 =\frac{\int \dd \phi\,\dd \phi'\,
 \langle \phi',\s|\,\rho_\s\,|\phi,\s\rangle
 \int_{\phi'}^{\phi} \D \varphi \,
 \Op (t_1)\cdots \Op (t_n)\,
 \e^{i{\cal A} [\varphi]}}
 {\int \dd \phi\,\dd \phi'\,
 \langle \phi',\s|\,\rho_\s\,|\phi,\s\rangle
 \int_{\phi'}^{\phi} \D \varphi \,
 \e^{i{\cal A} [\varphi]}}
\eeq
where $T_\Sp$ denotes time-ordering along the contour $\Sp$ and
$\Op (t) \equiv \Op [\varphi(t)]$. 
The classical action ${\cal A}$ is the time integral of the appropriate 
Lagrangian along the contour:
\beq
 {\cal A} [\varphi] \equiv \int_\Sp dt \,L [\varphi (t)]\,.
\eeq
In the above expression, $\int_{\phi'}^{\phi} \D \varphi$ denotes the 
integral over classical paths along the contour of Fig.\ \ref{fig:Spath}, 
with fixed boundaries at $t=0$ corresponding to the field configurations
$\phi'$ and $\phi$ respectively. The remaining integrals 
$\int \dd \phi\,\dd \phi'\,\langle \phi',\s|\,\rho_\s\,|\phi,\s\rangle$
represent the appropriate average over possible initial field configurations.
Note that the above formulation along the $\Sp$--path automatically includes 
the description of ``usual'' correlation functions with only positive time 
arguments. Note also that because of the $\Sp$--shape of the closed time 
path, positive times are always ``later'' (along the path) than negative 
times. Therefore, the time-ordering $T_\Sp$ always bring operators with 
negative-time arguments on the right of those with positive-time arguments. 

To illustrate this formalism, let us consider the time-evolution 
equation for the in-medium propagator 
of the fundamental field $\varphi$:\footnote{Notice that, throughout 
this section, we are focusing our attention on the internal dynamics of 
the emitting system and the propagator $G$ should not be confused with 
the free photon propagator (\ref{Fgauge}) introduced in section~\ref{sec:emrad}.}
\beq
\label{prop}
 G(x,y)=\langle T_\Sp\{\varphi_\s (x)\,\bar{\varphi}_\s (y)\}\rangle_\s\,.
\eeq
where we have reintroduced spatial variables: $x\equiv(x^0,\bx)$ and
similarly for $y$. Here, we use a unified notation 
where $\bar{\varphi}\equiv \varphi^\dagger$ for bosonic fields and 
$\bar{\varphi}\equiv \varphi^\dagger\gamma_0$ for fermionic fields.
The Schwinger-Dyson equation for the propagator (\ref{prop}) reads:
\beq
\label{SDeq}
 G^{-1} (x,y) = G_0^{-1} (x,y) - \Sigma(x,y)
\eeq
where $\Sigma(x,y)$ is the corresponding self-energy and 
\beq
 i G^{-1}_0(x,y) \equiv 
 \frac{\delta^2 {\cal A}[\varphi]}{\delta\varphi(x)\delta\bar{\varphi}(y)}
\eeq
is the inverse classical propagator. The latter usually takes the form 
of a local differential operator:
\beq
 i G^{-1}_0(x,y) = \D (x) \, \delta_\Sp(x-y)
\eeq
where $\delta_\Sp(x-y)\equiv \delta_\Sp (x^0-y^0)\,\delta^{(3)} (\bx-\by)$.
It corresponds, for instance, to the Klein-Gordon operator 
$\D (x) = -(\square_x+m_b^2)$ in the case of a scalar field
with bare mass $m_b$, or to the Dirac operator $\D (x) = 
(i\partial \,\slash\!_x-M)$ for a fermionic field with 
mass $M$.
 
As usual, one introduces the various components of two-point functions 
as (upper/lower sign corresponds to boson/fermion two-point function):
\beq
 G(x,y)=\Theta_\Sp (x^0-y^0)\,G^>(x,y) \pm \Theta_\Sp (y^0-x^0)\,G^<(x,y)\, ,
\eeq
as well as the spectral component:
\beq
\label{spectralpm}
 G_\rho(x,y)=i\Big( G^>(x,y) \mp G^<(x,y) \Big)\,,
\eeq
and similarly for the self energy $\Sigma (x,y)$.
The Schwinger-Dyson equation (\ref{SDeq}) can be rewritten in the form
of time-evolution equations for independent components $G^>$ and $G^<$ 
by means of standard manipulations (see e.g.\ \cite{Berges:2001fi,Berges:2002wr}). 
One obtains\footnote{In deriving this equation, one makes use of the 
equal-time (anti-)commutation relations of field operators, which 
imply the following relations: 
$G_\rho(x,y)|_{x^0=y^0}=0$ and $\partial_{x^0}G_\rho(x,y)|_{x^0=y^0}= 
\delta^{(3)}(\bx-\by)$ for bosons, and 
$G_\rho(x,y)|_{x^0=y^0}=i\gamma_0\delta^{(3)}(\bx-\by)$ for fermions.}
\beq
\label{evol}
 \D (x)\,G^{>,<} (x,y)=
 \int_0^{x^0} \!\dd u\,\,\Sigma_\rho (x,u)\,G^{>,<} (u,y)
 -\int_0^{y^0} \!\dd u\,\,\Sigma^{>,<} (x,u)\,G_\rho(u,y)\, .
\eeq
with the notation $\int_0^{x^0}\!\dd u\equiv\int_0^{x^0} \!du^0
\int d^3u$.
Note that these equations have the same form in the various cases 
where $x^0$ and $y^0$ are either both positive, both negative, or have 
opposite signs.\footnote{For practical calculations (see 
e.g.~\cite{Berges:2001fi,Berges:2002wr}), 
it often proves useful to employ the so-called statistical two-point 
function, defined as $G_F (x,y)=\frac{1}{2}\Big(G^>(x,y)\pm G^<(x,y)\Big)$
(and similarly for the self energy), together with the spectral function 
(\ref{spectralpm}) as independent variables. We mention that the 
corresponding evolution equations have a similar form as Eq.\ (\ref{evol}), 
and can actually be obtained from the latter by replacing, on both
side of the equality, $G^{>,<}\rightarrow G_F$ and 
$\Sigma^{>,<}\rightarrow \Sigma_F$ for the statistical function 
and $G^{>,<}\rightarrow G_\rho$ and $\Sigma^{>,<}\rightarrow \Sigma_\rho$ 
for the spectral function.} 
Moreover, we observe that the evolution equation for correlation
functions with $x^0$ and $y^0$ both positive (negative) only involve 
unequal-time two-point functions with positive (negative) time 
arguments. In contrast, in the case where $x^0$ and $y^0$ have 
opposite signs, the memory integrals on the RHS of the evolution 
equation (\ref{evol}) involve `positive-positive', `negative-negative' 
as well as `positive-negative' unequal-time two-point functions.
For a given expression of the self energy, one can, in principle, solve 
these time-evolution equations starting from given out-of-equilibrium 
initial conditions, e.g.\ along the lines of Refs.\ 
\cite{Berges:2001fi,Berges:2002wr}. The main advantage of the present 
formalism is that the initial conditions are to be specified at the 
physical initial time $t=0$. One may worry about the fact 
that the negative-time integration, e.g. in Eq.\ (\ref{gammanptime}) 
above, formally extends to the infinite past. In practice, however, 
unequal-time correlation functions are damped for large time differences, 
due to non-linear scattering effects and one therefore expects the 
negative-time integration to be effectively restricted to a finite 
range. 

Important progress have been made in recent years in describing the
far-from-equi\-li\-bri\-um dynamics of quantum fields (for a recent
review, see~\cite{Berges:2003pc}). Practicable approximation schemes 
may be based on the so-called two-particle-irreducible (2PI) effective 
action formalism~\cite{Berges:2000ur}, or on related truncations of 
Schwinger-Dyson equations~\cite{Mihaila:2000sr}. In particular, these 
methods have proven a powerful tool for explicit, first principle 
calculations of the time evolution of equal as well as unequal-time 
correlation functions. Moreover, they are particularly well-suited to 
describe the damping effects discussed above. For instance, the latter 
can be systematically taken into account in a 
loop~\cite{Berges:2000ur,Aarts:2001qa} 
or $1/N$~\cite{Berges:2001fi,Aarts:2002dj} expansion of the 2PI 
effective action beyond leading-order (mean-field) approximations. 
These methods might easily be combined with the present 
$\Sp$--path formalism in order to study electromagnetic radiation 
in genuine non-equilibrium situations.

\subsection{The current-current correlator from the 2PI effective action}

\begin{figure}[t]
 \begin{center}
 \epsfig{file=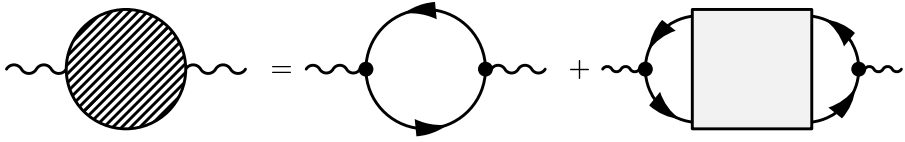,width=12.cm}
 \end{center}
 \caption{\label{fig:fullpolar}
 The complete non-equilibrium photon polarization tensor at lowest order
 in $e$ in terms of the full in-medium two-point function $D_\psi$ (lines) 
 and amputated four-point function $\Gamma^{(4)}$ (grey box) in a general theory 
 with charged fermionic fields (for a theory with charged scalar fields 
 there would be an additional tadpole-like contribution).}
\end{figure}

We end this paper by discussing how the photon polarization tensor can be
obtained from the 2PI effective action. In general, the electromagnetic 
current operator is a bilinear in the charged fields describing the 
emitting system and the current-current correlator (\ref{polar}) therefore 
involves the four-point function of the theory. The 2PI generating 
functional automatically provides infinite ladder-type resummations 
for the latter~\cite{McKay:1989rk}.
To illustrate this point we consider a fermionic field theory with
electromagnetic current given by (in the 
Heisenberg picture):\footnote{Here, $\psi$ and $\bar{\psi}$
denotes the fundamental fields of the emitting system and should not
be confused with the leptonic fields of section~\ref{sec:emrad}.}
\beq
 J_\s^\mu(x)=e\,\bar{\psi}_\s (x)\gamma^\mu \psi_\s (x)\,.
\eeq
The exact current-current correlator is graphically 
represented on Fig.\ \ref{fig:fullpolar} and can be expressed as follows:
\bea
\label{polarexp}
 \Pi^{\mu\nu} (x,y) &=&
 \Big<T_\Sp\Big\{J_s^\mu(x)\,J_s^\nu(y)\Big\}\Big>_s\nonumber\\
 &=& -e^2\,\tr[\gamma^\mu D_\psi(x,y)\gamma^\nu D_\psi(y,x)]
 -e^2\,{\cal V}^\mu(x): \Gamma^{(4)}:{\cal V}^\nu (y) \,,
\eea
where the first term on the RHS is the one-loop contribution, with
\beq
 D_\psi(x,x')=\Big<T_\Sp\Big\{\psi_s (x)\,\bar{\psi}_s (x')\Big\}\Big>_s
\eeq
the full in-medium fermion propagator, and the second term is the 
contribution from the amputated four-point function $\Gamma^{(4)}$ (see 
Eq.\ (\ref{fourpoint}) below), which is represented by the grey box in 
Fig.\ \ref{fig:fullpolar}. In Eq.\ (\ref{polarexp}) above, we have 
defined the ``vertex'' function
\beq
\label{vertex}
 {\cal V}^\mu_{\alpha\beta} (x;u,v)\equiv 
 \Big(D_\psi (u,x)\,\gamma^\mu\, D_\psi(x,v)\Big)_{\alpha\beta}\,,
\eeq
where $\alpha$ and $\beta$ denote Dirac indices,
and we have introduced the notation
\beq
 {\cal V}^\mu(x): \Gamma^{(4)}:{\cal V}^\nu (y)
 \equiv \int_\Sp \dd u\,\dd \bar{u}\,\dd v\,\dd \bar{v}\,\,
 {\cal V}^\mu_{\alpha\bar{\alpha}}(x;u,\bar{u})\,
 \Gamma^{(4)}_{\alpha\bar{\alpha};\beta\bar{\beta}} (u,\bar{u};v,\bar{v})\,
 {\cal V}^\nu_{\beta\bar{\beta}} (y;v,\bar{v})\,,
\eeq
where a sum over repeated Dirac indices is implied.\footnote{Note that 
whenever Dirac indices are involved, they are ordered and summed over in 
the same way as space-time variables. In the following, we omit them for 
simplicity.} Here, $\Gamma^{(4)}$ denotes the connected four-point function 
amputated from its external legs:
\beq
\label{fourpoint}
 \Big<T_\Sp\Big\{\psi_\s (x_1)\bar{\psi}_\s (x_2)
 \psi_\s (x_3)\bar{\psi}_\s (x_4)\Big\}\Big>_\s
 =-\Delta (x_1,x_2):\Gamma^{(4)}:\Delta (x_3,x_4)\,,
\eeq
where we introduced the ``two-particle'' function
\beq
 \Delta (u,\bar{u};v,\bar{v})=D_\psi(u,\bar{v})\,D_\psi(v,\bar{u})
\eeq
and where we employed a similar notation as before:
\beq
 \Delta (x_1,x_2):\Gamma^{(4)}:\Delta (x_3,x_4) \equiv 
 \int_\Sp \dd u\,\dd \bar{u}\,\dd v\,\dd \bar{v}\,\,
 \Delta (x_1,x_2;u,\bar{u})\,\Gamma^{(4)} (u,\bar{u};v,\bar{v})\,
 \Delta (v,\bar{v},x_3,x_4)\,.
\eeq

In order to illustrate how the 2PI effective action formalism might be 
used to compute the amputated four-point function, we consider the 
example of a purely fermionic theory with classical action:
\beq
 {\cal A} [\psi,\bar{\psi}] = \int {\rm d}^4 x \Big(
 \bar{\psi}(x) [ i \partial\,\slash - M ] \psi(x) 
 + V(\psi,\bar{\psi}) \Big) \, 
\eeq
For the relevant case of a vanishing fermionic ``background'' field,
the corresponding 2PI effective action $\Gamma$ can be written as~\cite{Cornwall:1974vz,
Berges:2002wr}:
\beq
 \Gamma[D_\psi] = 
 -i \Tr\ln D_\psi^{-1} -i \Tr\, D_0^{-1} D_\psi
 + \Gamma_2[D_\psi] + {\rm const} \, ,
\eeq
where the inverse classical propagator is given by
\beq
 i D_0^{-1} (x,y) 
 = (i \partial\,\slash\!_x - M )\, \delta_\Sp(x-y)\, .
\eeq  
The exact expression for the functional $\Gamma_2[D_\psi]$ contains all 
2PI diagrams with vertices described by $V(\psi,\bar{\psi})$ and propagator 
lines associated to the full fermion propagator $D_\psi$. The functional
trace $\Tr$ includes an integration over the closed time path $\Sp$ as 
well as integration over spatial coordinates and summation over Dirac 
indices. In absence of external sources, the equation of motion for $D_\psi$ 
is obtained by minimizing the effective action:
\beq
 \frac{\delta\Gamma[D_\psi]}{\delta D_\psi(x,y)} = 0\,,
\eeq
which leads to the following Schwinger-Dyson equation for the two-point
function:
\beq
 D_\psi^{-1} (x,y) = D_0^{-1} (x,y) - \Sigma_\psi (x,y;D_\psi) \, ,
\eeq
with proper self-energy given by:
\beq 
 \Sigma_\psi(x,y;D_\psi) =
 -i\frac{\delta \Gamma_2[D_\psi]}{\delta D_\psi(y,x)} \, .
\eeq  

\begin{figure}[t]
 \begin{center}
 \epsfig{file=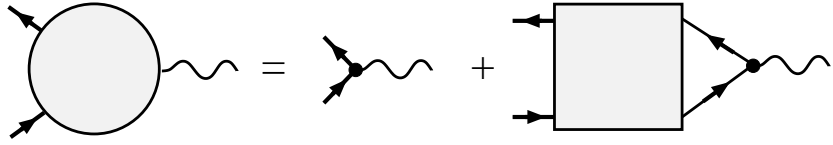,width=11.cm}
 \end{center}
 \caption{\label{fig:BSdef} Definition of the resummed vertex $V^\mu$ 
   (grey circle) in terms of the bare vertex ($\gamma^\mu$) and the amputated 
   four-point function $\Gamma^{(4)}$ (grey square).}
\end{figure}

\begin{figure}[t]
 \begin{center}
 \epsfig{file=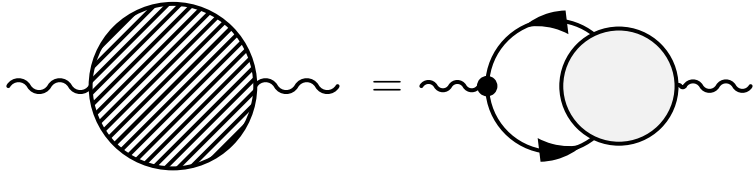,width=10.cm}
 \end{center}
 \caption{\label{fig:fullpolar2} The non-equilibrium photon polarization tensor
   $\Pi^{\mu\nu}$ expressed in terms of the resummed vertex $V^\mu$ (grey circle).}   
\end{figure}

\begin{figure}[h]
 \begin{center}
 \epsfig{file=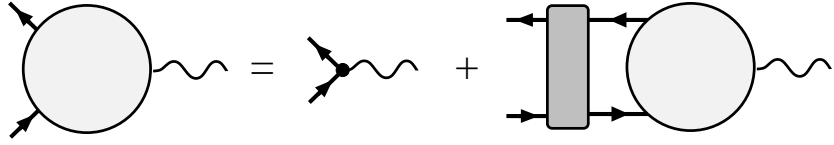,width=11.cm}
 \end{center}
 \caption{\label{fig:BSeq2} The integral equation satisfied by the resummed vertex 
   $V^\mu$ (grey circle). The dark grey box represents the kernel $\Lambda$. 
   This equation resums an infinite series of ladder graphs with rungs corresponding 
   to $\Lambda$.}
\end{figure}

Higher correlation functions can be obtained from appropriate functional 
derivatives of the 2PI effective action, which generically leads to 
integral equations. For instance, one can show that the amputated four-point 
function satisfies the following Bethe-Salpeter equation~\cite{McKay:1989rk} 
(space-time variables and Dirac indices are implicit):\footnote{This equation 
plays an important role for the issue of renormalization in the 2PI 
formalism~\cite{vanHees:2001ik,Blaizot:2003br}.}
\beq
\label{BSeq}
 \Gamma^{(4)}=\Lambda+\Lambda:\Delta: \Gamma^{(4)}
\eeq
with kernel given by:
\beq
\label{BSkernel}
 \Lambda (u,v;\bar{u},\bar{v})=
 -i\frac{\delta^2 \Gamma_2 [D_\psi]}
 {\delta D_\psi (u,v)\,\delta D_\psi (\bar{u},\bar{v})}\,.
\eeq
\begin{figure}[t]
 \begin{center}
 \epsfig{file=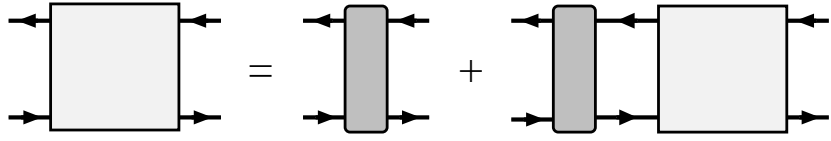,width=11.cm}
 \end{center}
 \caption{\label{fig:BSeq}
 The Bethe-Salpeter equation for the 
 four-point function $\Gamma^{(4)}$ (grey box) in a theory with quartic 
 interactions. The lines represent the full propagator of the theory ($D_\psi$) 
 and the dark grey box corresponds to the kernel $\Lambda$ of this integral 
 equation (see text).}
\end{figure}
Equation (\ref{BSeq}) is represented graphically on Fig.\ \ref{fig:BSeq}, 
where the light and dark grey boxes represent the four-point function 
$\Gamma^{(4)}$ and the kernel $\Lambda$ respectively.
This integral equation generates an infinite resummation of ladder graphs with 
rungs given by the kernel (\ref{BSkernel}). Thus we see that a {\em finite}-order 
truncation of the 2PI effective action in a given expansion scheme automatically 
leads to an {\em infinite} (ladder-type) resummation for the four-point function 
through Eq.\ (\ref{BSeq}).\footnote{Similar resummations in the 2PI formalism 
have been recently exploited for the computation of transport coefficient 
in $O(N)$ scalar theories \cite{Aarts:2003bk,Aarts:2004sd}.} These resummations 
correspond to the physics of multiple scatterings and actually generalize the 
gluon ladder resummation recently employed in the context of equilibrium QCD at 
high temperature to describe the LPM effect \cite{Arnold:2001ba,Aurenche:2002wq}. 
To make this last point more transparent, let us introduce the following 
resummed vertex function (see Fig.\ \ref{fig:BSdef}):\footnote{With the 
present notation, we have: $\Gamma^{(4)}(u,v):{\cal V}^\mu(x)\equiv\int_\Sp 
\dd{\bar u}\,\dd{\bar v}\,\Gamma^{(4)}(u,v;{\bar u},{\bar v}){\cal V}^\mu({\bar u},
{\bar v};x)$.}
\beq
\label{BSdef}
 V^\mu(x;u,v)\equiv\gamma^\mu\,\delta_\Sp(x-u)\delta_\Sp(x-v)
 +\Gamma^{(4)}(u,v):{\cal V}^\mu(x)\,,
\eeq
in terms of which the photon polarization tensor can be rewritten 
(see Fig.\ \ref{fig:fullpolar2}):
\beq
\label{polar2}
  \Pi^{\mu\nu} (x,y) = -e^2\,{\cal V}^\mu(x):V^\nu(y)\,.
\eeq
Using equation (\ref{BSeq}), it is now easy to show
that the resummed vertex (\ref{BSdef}) satisfies the following integral
equation (space-time variables and Dirac indices are implicit):\footnote{
Using Eq.\ (\ref{BSeq}) and the definition (\ref{vertex}), we have:
\bea
 V^\mu&=&\gamma^\mu+\Gamma^{(4)}:{\cal V}^\mu\nonumber\\
 &=&\gamma^\mu+\Lambda:{\cal V}^\mu+
 \Lambda:\Delta:\Gamma^{(4)}:{\cal V}^\mu\nonumber\\
 &=&\gamma^\mu+\Lambda:\Delta:\gamma^\mu+
 \Lambda:\Delta:\Gamma^{(4)}:{\cal V}^\mu\nonumber\\
 &=&\gamma^\mu+\Lambda:\Delta:V^\mu\,.\nonumber
\eea
} 
\beq
 V^\mu=\gamma^\mu+\Lambda:\Delta:V^\mu
\eeq
which is graphically depicted on Fig.\ \ref{fig:BSeq2}. This, together with
Eq.\ (\ref{polar2}) for the photon polarization tensor is a direct generalization
of the type of integral equation introduced in Refs.\ \cite{Arnold:2001ba,Aurenche:2002wq} 
to account for the LPM effect in equilibrium QCD. One can easily convince oneself 
that the resummation of one-gluon rungs needed there would simply correspond to a 
two-loop approximation of the corresponding 2PI effective action. It is remarkable 
that the infinite set of contributions needed to obtain the full photon and 
dilepton production rates at leading order in $\alpha_s$ actually correspond to a
finite, $\Op(\alpha_s)$, truncation of the 2PI effective action.
\footnote{For recent discussions of gauge theories in 
the context of generalized effective actions, see \cite{Arrizabalaga:2002hn,
Carrington:2003ut,Berges:2004pu}.} 

\acknowledgments
I would like to thank R. Baier, F. Gelis and D. Schiff for interesting
and stimulating discussions. I thank J.~Berges for fruitful collaboration
on related topics.

\end{document}